\documentclass[pra,aps,showpacs,groupauthors,footinbib,twocolumn]{revtex4-1}

\usepackage{amssymb}
\usepackage[english]{babel}
\usepackage[ansinew]{inputenc}
\usepackage{amsfonts}
\usepackage{amsmath}
\usepackage{graphicx}
\usepackage{bm}
\usepackage{color}

\newcommand{\ket}[1]{|{#1}\rangle}
\newcommand{\bra}[1]{\langle{#1}|}

\newcommand{\toni}[1]{\textcolor{black}{#1}}
\newcommand{\an}[1]{\textcolor{black}{#1}}
\newcommand{\ann}[1]{\textcolor{black}{#1}}
\newcommand{\lm}[1]{\textcolor{black}{#1}}
\newcommand{\gan}[1]{\textcolor{black}{#1}}
\newcommand{\lmn}[1]{\textcolor{black}{#1}}

\newcommand{\tunnu}[1]{\textcolor{black}{#1}}

\begin{document}


\title{\an{Quantum computing implementations with neutral particles}}


\author{Antonio Negretti$^{1,2}$}
\author{Philipp Treutlein$^3$}
\author{Tommaso Calarco$^1$}

\affiliation{
1. Institute for Quantum Information Processing, University of Ulm, Albert-Einstein-Allee 11, D-89069 Ulm,
Germany.\\
\gan{2. Lundbeck Foundation Theoretical Center for
Quantum System Research, Department of Physics and Astronomy,
University of Aarhus, DK-8000 Aarhus C, Denmark}\\
3. Departement Physik, Universit\"at Basel, Klingelbergstrasse 82, CH-4056 Basel, Switzerland.}

\date{\today}


\begin{abstract}
\an{We review quantum information processing with cold neutral particles, that is, atoms or polar molecules. 
First, we analyze the best suited degrees of freedom of these particles for storing quantum information, and then we 
discuss both single- and two-qubit gate implementations. We focus our discussion mainly on collisional quantum gates, 
which are best suited for atom-chip-like devices, as well as on gate proposals conceived for optical lattices. 
Additionally, we analyze schemes both for cold atoms confined in optical cavities and hybrid approaches to 
entanglement generation, and we show how optimal control theory might be a powerful tool to enhance the 
speed up of the gate operations as well as to achieve high fidelities required for fault tolerant quantum computation.}
\end{abstract}




\maketitle


\section{Introduction}
\label{sec:intro}

Since the 1990s, when groundbreaking algorithms based on the laws of quantum 
mechanics for solving classically intractable computational problems were found, 
quantum information science has rapidly grown with the promise to build up a 
quantum computer. Similar to nowadays ``classical" computers, quantum hardware 
consists of a memory and a processor. The former stores the information, the latter, 
with a set of gates, processes the information. 

The concept of gate is fundamental in quantum computation \cite{QIP:Nielsen00}, 
and therefore we first consider its classical analogue. A gate on a classical computer, 
which implements a Boolean function, is a device that accomplishes a well-defined 
operation on one or more bits. For instance, CMOS transistors realize the logical 
NOT operation. Instead, a quantum gate performs a unitary transformation on the linear 
space of quantum bits (\emph{qubits}). Thus, a quantum gate is the time propagator 
generated by a given Hamiltonian; control by external fields, according to 
the Hamiltonian structure, allows to perform desired transformations on the qubit 
wave function. Again in the 90s it has been showed that a general $N$-qubit gate can be 
decomposed into $O(N^2)$ one- and two-qubit gates. As a consequence, most of the 
schemes for quantum gates concern the implementation of one- and two-qubit operations.

\an{Neutral particles such as cold atoms and polar molecules are excellent candidates 
for quantum information processing (QIP) implementations. Indeed, $(i)$ they have exquisite 
coherence properties; $(ii)$ their coherent evolution can be accurately controlled in tailored 
micropotentials; and $(iii)$ they present the exciting perspective of interfacing their qubit degrees of freedom 
with solid-state systems.}

\begin{figure}[tb]
\includegraphics[width=0.48\textwidth]{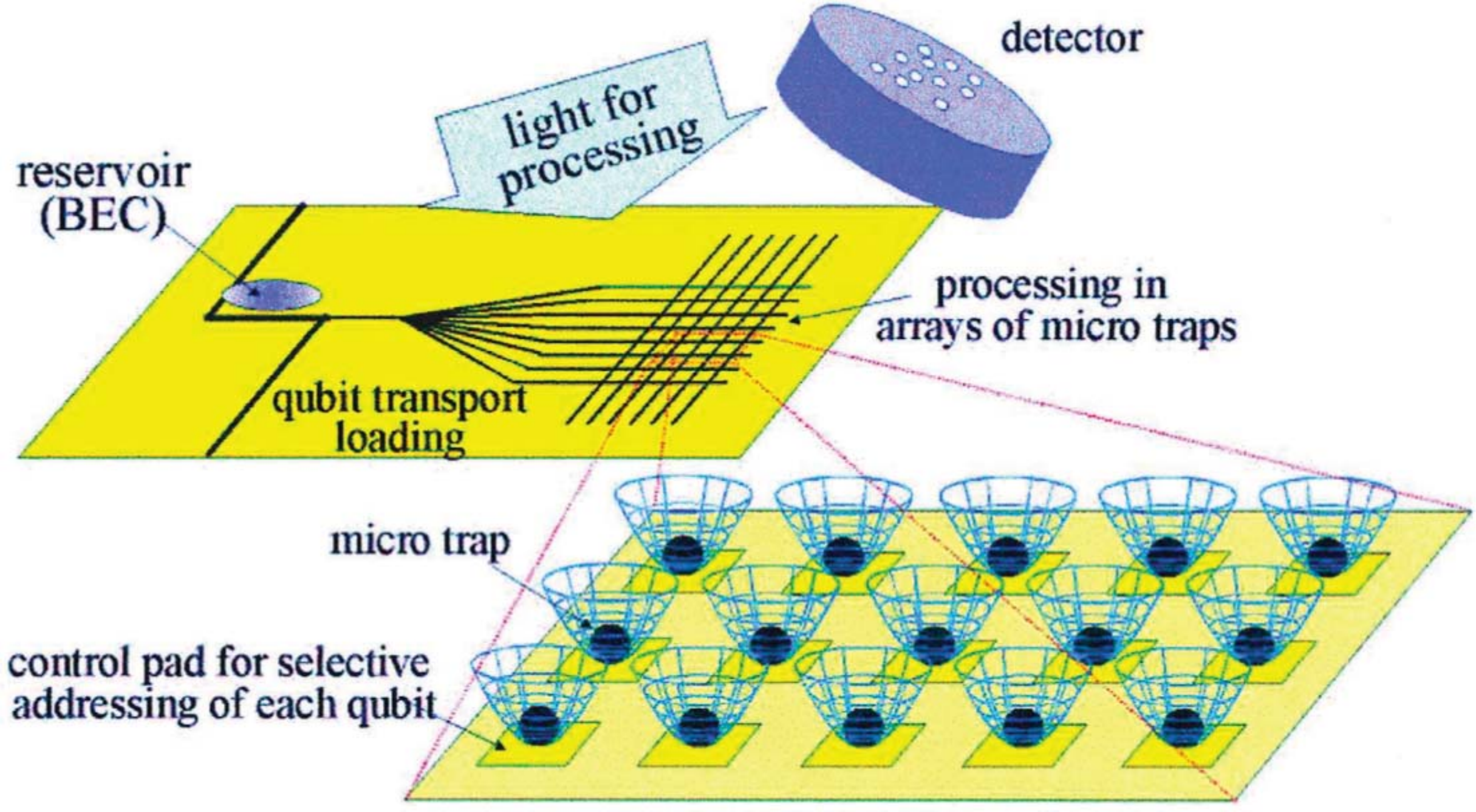}
\caption{Schematic illustration of an atom chip quantum processor, 
adapted from Ref.~\cite{QIP:Schmiedmayer02}.
}
\label{QIP:fig:quantumprocessor}
\end{figure}

\an{There are two main paradigms for a quantum hardware based on neutral particles in surface traps: the first 
one is a microfabricated device called atom chip, whereas the second one relies on the combination of 
both polar molecules and superconducting circuits. In figure~\ref{QIP:fig:quantumprocessor} 
the general idea of an atom chip quantum processor is illustrated.} It includes a reservoir of cold atoms, preferably 
in their motional ground state, and in a well-defined internal state. An ideal starting 
point for this is a Bose-Einstein condensate (BEC) in a chip trap. From there the atoms are transported using guides 
or moving potentials to a large array of processing sites. Either single atoms, or small 
ensembles of atoms, are then loaded into the qubit traps. Each qubit site can be addressed 
individually.  Microfabricated wires and electrodes located close to the individual sites can 
be used for site-selective manipulations such as single-qubit gates. For two-qubit gates, 
interactions between adjacent sites are induced. For readout, micro-optics can be used to 
focus lasers onto each site separately, or the whole processor can be illuminated and single 
qubits are addressed by shifting them in and out of resonance using local electric or magnetic 
fields. 

\begin{figure}[tb]
\includegraphics[width=0.45\textwidth]{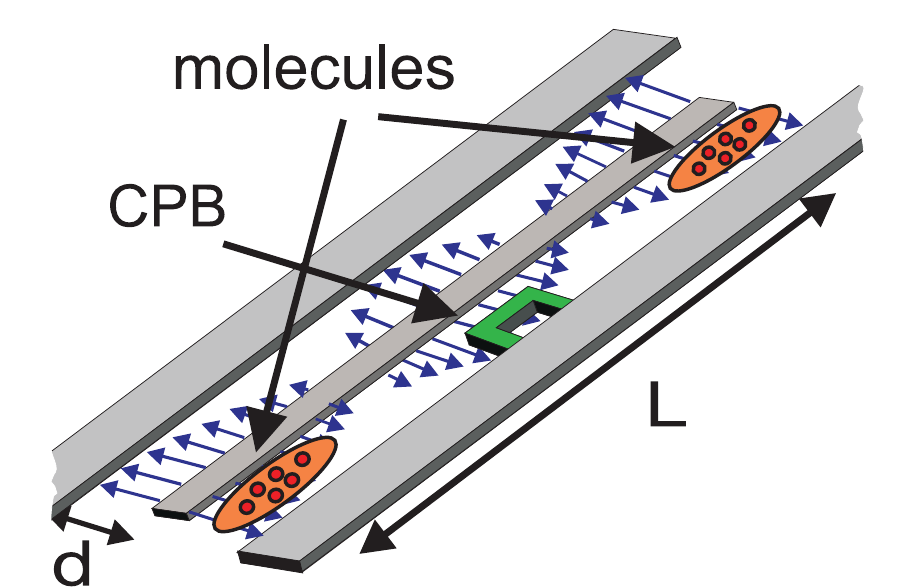}
\caption{\an{Hybrid quantum processor based on the combination of ensembles of polar molecules, 
superconducting transmission lines and Cooper-pair boxes (CPB), adapted from Ref.~\cite{QIP:Rabl06} . 
Copyright (2006) by The American Physical Society.}
}
\label{QIP:fig:Hquantumprocessor}
\end{figure}

\an{
The other model of quantum computation is illustrated in Fig.~\ref{QIP:fig:Hquantumprocessor}. 
Contrarily to the former, such a scheme employs molecular ensembles to store the quantum 
information, whereas the superconducting solid-state circuit is utilized for processing it. This 
quantum hardware paradigm brings the best features of the atomic, molecular, and 
solid-state systems together: the excellent 
coherence times of atoms and molecules, as previously outlined, and the extremely short 
manipulation times of solid-state qubits. The basic idea behind the processor displayed in 
Fig.~\ref{QIP:fig:Hquantumprocessor} is that the transmission line plays the role of a ``quantum bus", 
that is, each time quantum information has to be processed through a sequence of quantum 
gates, the molecular qubits are transferred to their solid-state counterparts (e.g., Cooper-pair 
boxes). The transfer is mediated by (microwave) photons in the transmission line which are 
almost resonant with both kinds of qubits. In this scenario the superconducting waveguide behaves 
as a single mode resonator, analogously to the situation encountered in quantum optics with optical 
cavities. The readout can be performed by measuring the phase or amplitude of the transmitted
radiation of a microwave driving field in the superconducting waveguide.
}

\toni{
\an{Even though such visions have not yet been completely 
realized experimentally, these concepts of a quantum processor are currently 
pursued in several laboratories and such ideas are becoming closer and closer to reality, 
as this special issue is reporting on.} Beside this, we would also like to underscore \tunnu{(a 
point that is usually not emphasized)}, that when 
quantum memories are discussed, such as the one depicted in Fig.~\ref{QIP:fig:quantumprocessor} 
(lower layer), reference is usually made to the QRAM, that is, the \an{quantum} analogue of the RAM 
(random-access memory) of classical computers. Indeed, \tunnu{most} of the proposed 
physical implementations for storing a bit of quantum information have coherence times 
$\lesssim$ 1 s.} \tunnu{Only for (single) trapped ions coherence times of minutes or even hours 
have been experimentally observed, and recently, but for an ensemble of a few thousand
neutral $^{87}$Rb atoms, a coherence time of almost a minute has been reported  \cite{QIP:Deutsch10}.
The long-lived memory, that is, up to several 
years of storage time without leakage of information, will still be a ``classical" 
device. Such a memory has to communicate (be interfaced) with the quantum processor, 
where the information is processed. }Given the fact that the long-lived memory is typically 
a solid-state device and given the intriguing perspective to perform experiments 
at the interface of quantum (atom) optics and solid-state physics, \an{atom-chip-like devices and 
superconducting electrical circuits offer an excellent platform 
to realize such quantum hardware paradigms.}

A fully developed architecture for quantum information processing  consists of the following elements 
\cite{QIP:diVincenzo00,QIP:Schmiedmayer02,QIP:Treutlein06a}: 
(i) qubit states with long coherence lifetime; 
(ii) qubit initialization;
(iii) single- and two-qubit gates;
(iv) qubit readout;
(v) interfaces to other systems.
\an{Here we will focus only on (i) and (iii) and discuss several computation schemes. In the 
present paper we will also discuss hybrid approaches, i.e. the combination of solid-state and 
atomic systems, such as the proposal of Ref.~\cite{QIP:Andre06}, but without entering into 
technical details} \gan{(see Sec.~\ref{sec:hybrid})}. \an{For further details we refer to Ref.
 \cite{QIP:ACbook11} and for an additional comparison among atomic, solid-state, linear 
 optical, and NMR based quantum computing implementation we refer to the excellent 
text book \cite{QIP:Chen06}.
}

\section{Suitable qubit states for QIP}
\label{QIP:sec:qubitstates}

Two potentially conflicting requirements have to be met by the qubit states 
$\{|0\rangle, |1\rangle \}$ chosen for QIP \an{with neutral particles}. On the one hand, 
both states have to couple to the electromagnetic fields which are used for 
trapping and manipulating, \an{for example,} the atoms.
On the other hand, high fidelity gate operations require a long coherence 
lifetime of superposition states $\alpha |0\rangle + \beta |1\rangle$, 
$(|\alpha |^2 + |\beta |^2 = 1)$, and thus the qubit has to be sufficiently 
robust against fluctuations of electromagnetic fields in realistic experimental 
situations. A peculiarity of chip-based traps is the presence of atom-surface 
interactions, which lead to additional loss and decoherence mechanisms which 
are not present in macroscopic traps. 
It is therefore important to investigate the coherence properties of the proposed 
qubit candidates close to the chip surface. In the following we discuss the different 
types of qubits which have been studied.

\subsection{Hyperfine qubits}
\label{QIP:sec:qubitstatesHyper}

An obvious qubit candidate are the long-lived \lmn{(electronic)} ground state hyperfine levels of the atoms. 
In most experiments, at least a part of the trapping potential is provided by static magnetic 
fields generated by wires or permanent magnet structures on chip. It is therefore 
desirable that both $|0\rangle$ and $|1\rangle$ are magnetically trappable. Long coherence 
lifetimes can be expected if states with equal magnetic moments are chosen, so that both 
states experience nearly identical trapping potentials in static magnetic traps, and the 
energy difference $h\nu_{10} = E_{|1\rangle}-E_{|0\rangle}$ between $|0\rangle$ 
and $|1\rangle$ is robust against magnetic field fluctuations.
These requirements are satisfied by the $|F=1,m_F=-1\rangle \equiv |0\rangle$ 
and $|F=2,m_F=+1\rangle \equiv |1\rangle$ hyperfine levels of the $5S_{1/2}$ 
ground state of $^{87}$Rb. 

In Ref.~\cite{QIP:Treutlein04}, the coherence properties of this qubit pair were studied experimentally 
on an atom chip. A coherence lifetime exceeding 1~s was measured at different atom-surface 
distances using Ramsey spectroscopy. These experiments confirm that this hyperfine qubit 
is very well suited for \an{atom-chip based QIP. It is therefore considered in several schemes 
for quantum gates and we shall describe gate implementations using this qubit 
in some detail later in the paper.}

Magnetic field insensitive hyperfine qubits were also experimentally studied in optical microtraps, 
created by an array of microlenses \cite{QIP:Lengwenus07}. Using spin-echo techniques, coherence 
times of $68$~ms were obtained, limited by spontaneous scattering of photons. Simultaneous Ramsey 
measurements in up to 16 microtraps were performed, demonstrating the scalability of this approach.

\lmn{
Finally, we also note that by using alkaline-earth(-like) atoms one can encode
qubits in nuclear spin states, decoupled from the electronic state in both the 
$^{1}S_0$ ground state and the very long-lived $^{3}P_0$ metastable state~\cite{Daley2008}.
}

\subsection{Motional qubits}

Qubits can also be encoded in the vibrational states of atoms in tight traps. This has been proposed 
both for optical \cite{QIP:Eckert02,QIP:Mompart03} and magnetic \cite{QIP:Cirone05,QIP:Charron06} 
microtraps. The computational basis states can be two vibrational levels in a single trap, e.g.\ the ground 
and first excited vibrational level \cite{QIP:Eckert02}. Alternatively, they can be defined by the presence 
of an atom in either the left or the right well of a double well potential \cite{QIP:Mompart03}. 
Initialization of the atoms in the lowest vibrational state of the trap with high fidelity is crucial in both 
of these schemes. 

Vibrational states are usually more delicate to handle and to detect than hyperfine states. An advantage, 
on the other hand, is that an internal-state independent interaction is sufficient for two-qubit gates and 
collisional loss can be reduced as the two interacting qubits are in the same internal state. The proposals 
of Refs.~\cite{QIP:Cirone05,QIP:Charron06} therefore consider a combination of hyperfine states for qubit 
storage and vibrational states for processing. Measurements of vibrational coherence near surfaces still 
have to be performed. The expected fundamental limits due to surface-induced decoherence, however, 
are comparable to those for hyperfine states.

\subsection{Rydberg state qubits}

Rydberg states are attractive for QIP because of their strong electric dipole moment \cite{QIP:Jaksch00,QIP:Mozley05}. 
The resulting dipole-dipole interaction between Rydberg atoms can be exploited for fast two-qubit quantum gates. 
Moreover, Rydberg qubits can be combined with long-lived ground state hyperfine qubits for information storage.
For the $n\sim 50$ Rydberg states of Rb, typical lifetimes are $\sim 100~\mu$s for low angular momentum states 
up to $\sim 30$~ms for circular states. In Ref.~\cite{QIP:Mozley05} it is proposed to enhance the lifetime of circular 
Rydberg states into the range of seconds by using a microstructured trap on a superconducting (atom) chip that 
simultaneously acts as a cavity with a microwave cut-off frequency high enough to inhibit spontaneous emission. 
Furthermore, it is shown that with the help of microwave state dressing, coherence lifetimes of similar magnitude 
could be achieved.

\subsection{Ensemble-based qubits}

A single qubit can be encoded in collective states of an ensemble of particles \cite{QIP:Lukin01,QIP:Brion07}. 
State $\ket{0}$ corresponds to all particles in the ground state, while in state $\ket{1}$, a single excitation is 
shared collectively by the whole ensemble. To isolate this two-level system, a blockade mechanism is required 
that prevents the creation of two or more excitations. The necessary nonlinearity can be provided by the 
dipole-dipole interaction of Rydberg atoms \cite{QIP:Lukin01,QIP:Brion07,QIP:Yan08}, or by coupling the 
ensemble via a cavity to a single saturable two-level system such as a Cooper pair box \cite{QIP:Rabl06,QIP:Verdu09}.
Ensemble qubits have the advantage that single-atom preparation is not required. Moreover, the Rabi 
frequency between the collective qubit states is enhanced by $\sqrt{N}$, where $N$ is the number of 
atoms in the ensemble. The decay rates are the same as for a single particle if the decay is dominated 
by non-collective processes such as atom loss. Ensemble qubits in chip traps have been considered for 
ground state atoms \cite{QIP:Verdu09}, Rydberg atoms \cite{QIP:Yan08}, polar molecules 
\cite{QIP:Rabl06,QIP:Tordrup08}, and electron spins \cite{QIP:Wesenberg09}. Recently, for the latter, 
an experiment demonstrates the storage and retrieval of up to 100 weak 10 GHz coherent excitations in such 
collective qubit states \cite{QIP:Wu10}.


\section{Single-qubit operations}
\label{QIP:sec:qubitrot}

Single-qubit gates are unitary transformations in the Hilbert space of a single qubit. The necessary degree 
of control can be experimentally demonstrated by driving high-contrast Rabi oscillations between the qubit states. 
Atom-chip based experiments have demonstrated such control with hyperfine qubit states. Coherent control of motional 
states has been demonstrated in the context of atom interferometry. 
Given that, in this section we only focus on hyperfine and motional qubit states. Beside this, the schemes based, 
for instance, on Rydberg states, utilize similar methods and (experimental) techniques for implementing single 
qubit rotations than those described here.

\subsection{Hyperfine qubits}
\label{QIP:sec:qubitrothyer}

Hyperfine qubit states can be coupled with oscillating microwave and/or radio frequency (rf) magnetic fields. 
High fidelity Rabi oscillations on the qubit transition \ann{$\ket{0}\equiv |F=1,m_F=-1\rangle \leftrightarrow \ket{1}\equiv |F=2,m_F=+1\rangle$} 
of $^{87}$Rb, whose first order Zeeman shift is approximately identical, have already been demonstrated 
experimentally on an atom chip~\cite{QIP:Treutlein04,QIP:Boehi09}. In this case, a two-photon transition with 
a microwave and an rf photon is involved. A two-photon Rabi frequency of a few kHz is easily achieved, so 
that single-qubit gates can be performed on a time scale of hundreds of microseconds (with a $\pi$-pulse 
fidelity of $>96$\%), three to four orders of magnitude faster than the relevant coherence lifetimes. 
Alternatively, hyperfine states can be coupled through a two-photon Raman transition driven by two laser 
beams \cite{QIP:Wineland98,QIP:Lengwenus07}. 

\an{We note that in atom-chip systems} the qubit driving fields can be generated by chip-based waveguides 
(for microwaves) or just simple wires (for rf). This results in a stable, well-controlled coupling with tailored 
polarization. All elements for qubit manipulation can thus be integrated on chip. Moreover, chip-based driving 
fields can have strong near-field gradients. This is advantageous as it allows individual addressing of spatially 
separated qubits. On the other hand, care has to be taken to avoid dephasing due to strong gradients across 
a single qubit.

\lm{The above discussion concerns implementations in atom-chip systems, but, as we shall discuss 
in Sec.~\ref{sec:optical}, optical lattices offer also an interesting platform for carrying out QIP protocols, 
especially because with them it is possible, simultaneously, to coherently manipulate millions of 
qubits~\cite{QIP:Mandel04}. The detection and manipulation of a single atom amongst an ensemble of $\sim 10^5$ atoms 
presents more difficulties than in an atom chip. Since the addressing occurs with a focused laser beam, the spacing 
between (nominally) identical lattices sites has to be larger than the optical diffraction limit, or, alternatively, arrays 
of independent single-atom traps sufficiently spaced can be used~\cite{QIP:Schrader04,QIP:Nelson07,QIP:Beugnon07}. 
The drawback of such approaches, however, is the need to use non-clock states, that is, field-sensitive states, thus 
with shorter coherence times, a requirement that is conflict with the third DiVincenzo criterion for quantum 
computing~\cite{QIP:diVincenzo00}, namely (i) in our introduction.}
\ann{
Recently, however, it has been shown, that in addition to the aforementioned qubit pair states, used to store the information, 
the field insensitive pair  $\ket{0^{\prime}}\equiv |F=2,m_F=0\rangle$ $\ket{1^{\prime}}\equiv |F=1,m_F=0\rangle$ 
of $^{87}$Rb (referred to as ``working" states) enables to obtain a fidelity for single qubit operations also larger than 
96\%~\cite{QIP:Lundblad09}. This can be achieved by mapping a superposition state of a storage qubit to the 
corresponding working states, because while the transitions $\ket{0}\leftrightarrow\ket{1}$, 
$\ket{0^{\prime}}\leftrightarrow\ket{1^{\prime}}$ are insensitive to external fields, the transitions $\ket{0}\leftrightarrow\ket{0^{\prime}}$, 
$\ket{1}\leftrightarrow\ket{1^{\prime}}$ are instead field sensitive. The application of a magnetic field gradient, 
which shifts the energy levels (Zeeman effect), allows to spectrally select a particular qubit from the 
quantum register. Field-sensitive transitions can then be used to transfer the storage-state superposition 
to the pair of working states for the selected qubit register with a fidelity of 99\%. At this point, an arbitrary qubit 
rotation can be performed on the selected qubit alone, since the working-state transition is off-resonant from the 
storage-state transition. Then, the inverted transfer process returns the selected qubit register to a new 
storage-state superposition. Such a technique is quite interesting and promising, because the spatially 
inhomogeneous external magnetic field used to map a storage-state superposition of the selected qubits 
to a working-state superposition, with a minimal perturbation of the unselected qubits, is not spatially localized. 
Hence, such a procedure makes the addressability of qubits less complicated while decoherence effects are 
substantially suppressed.  
}

\lmn{
Finally, we underscore that recently also single atom addressability in a two-dimensional geometry has been experimentally 
achieved, by using either a tightly focused laser beam together with a microwave field~\cite{Weitenberg2011} or by means of 
techniques from solid immersion microscopy~\cite{Bakr2010}. These results have demonstrated a sub-diffraction-limited 
resolution, significantly below the lattice spacing. 
}

\subsection{Motional qubits}

If the qubit is encoded in the ground and first excited vibrational levels of a single trap 
\cite{QIP:Eckert02,QIP:Cirone05,QIP:Charron06}, qubit rotations can be induced by 
driving a two-photon Raman transition between the states with two lasers. Such 
transitions between vibrational levels are routinely employed in ion trap QIP 
\cite{QIP:Wineland98} and have been demonstrated in optical dipole traps 
\cite{QIP:Morinaga99}. Similar experiments with neutral atoms in chip traps 
still have to be performed. As the vibrational levels have to be spectrally 
resolved, tight traps with large vibrational frequencies are required. On 
atom chips, sufficiently high vibrational frequencies of up to $\sim 1$~MHz 
are accessible. If the qubit basis states are the left and right states of a double 
well \cite{QIP:Mompart03}, single-qubit gates can be performed by adiabatically 
lowering the barrier between the two wells and allowing tunneling to take place. 
This has strong connections to atom interferometry. Chip-based atom interferometers 
demonstrating versatile coherent control of the motional state of BECs have been 
realized, see e.g.\ \cite{QIP:Wang05,QIP:Hofferberth06,QIP:Boehi09}.


\section{Conditional dynamics}
\label{QIP:sec:cond-dyn}

Two-qubit gates are the heart of a quantum processor, as they are required for the 
generation of entanglement between the qubits. 

Let us consider the dynamics of an arbitrary number of \an{particles} (no matter if 
charged or not) in a time- and state-dependent three-dimensional 
trapping potential $V_k({\bf r}, t)$ [${\bf r} = (x,y,z)$] governed by the 
Hamiltonian operator \cite{QIP:Calarco00a,QIP:Calarco00b}

\begin{align}
\label{eq:many-bodyH}
\hat H(t) &= \sum_{k=0}^1 \int{\rm }d{\bf r}\,\hat\Psi_k^{\dagger}({\bf r})
\left[-\frac{\hbar^2}{2 m}\nabla^2 
+ V_k({\bf r}, t)\right]\hat\Psi_k({\bf r})\nonumber\\
& \phantom{=}\!\!\!\!+\sum_{k,\ell=0}^1\frac{1}{2}\int{\rm }d{\bf r}d{\bf
  r}^{\prime}
\,\hat\Psi_k^{\dagger}({\bf r})
\hat\Psi_{\ell}^{\dagger}({\bf r}^{\prime})U_{k\ell}({\bf r},{\bf r}^{\prime})
\hat\Psi_{\ell}({\bf r}^{\prime})
\hat\Psi_k({\bf r}).
\end{align}
Here $m$ is the \an{mass of the particle}, $\hat\Psi_k^{\dagger}({\bf r}),\hat
\Psi_k({\bf r})$ are field creation and annihilation operators for 
the logic state $\ket{k}$, and $U_{k\ell}({\bf r},{\bf r}^{\prime})$ is the 
\an{two-body} interaction potential for the qubit states 
$\ket{k}$ and $\ket{\ell}$, with $k,\ell = 0,1$. 
Our goal is the realization of a two-qubit gate with two \an{neutral particles (e.g., atoms)}, each of them 
carrying a qubit of information usually encoded in an extra degree of
freedom (e.g., a pair of hyperfine states) other than their motional state. In this
specific case, the full many-body problem described by the Hamiltonian 
(\ref{eq:many-bodyH}) can be reduced to a Schr\"odinger equation for two 
trapped particles and this will be assumed in the following. 

The quantum gate we aim to implement is a phase gate having the
following truth table: 
$\ket{\epsilon_1}\ket{\epsilon_2} \rightarrow 
e^{i\phi_g\epsilon_1\epsilon_2} \ket{\epsilon_1}\ket{\epsilon_2}$, 
where $\ket{\epsilon_1},\ket{\epsilon_2}$ are the logic qubit states with 
$\epsilon_{1,2} = 0,1$. When the phase $\phi_g$ 
takes on the value of $\pi$, the combination of a phase gate with two Hadamard 
gates yields a controlled-NOT gate. In this respect it is an important quantum
gate. Since it requires only to produce a phase shift for the state
$\ket{1}\ket{1}$ such a gate has become of interest, because it requires a
state-dependent interaction that is relatively straightforward to realize physically.

Let us explain the basic principle to obtain a conditional phase shift $\phi_g$
 when two \an{neutral particles} are trapped in a microscopic potential. Initially, at $t=0$, we assume
that the two particles are in the respective ground states of the trapping 
potential and that their wave functions are well separated from each other so 
that their overlap is negligible. At times $0<t<\tau_g$ the potential wells are 
changed in such a way that the \an{particle} wave functions are displaced differently 
depending on their logical state $\ket{k}$ and a state-dependent wave function overlap results. 

The particles interact for a time $\tau_g$, the gate operation time, and at $t=\tau_g$ the 
initial situation is restored. With this approach we get state dependent 
phase shifts of two kinds: a purely kinematic one, $\phi_k + \phi_{\ell}$, 
due to the single particle motion in the trapping potential; and an interaction 
phase, $\phi_{k\ell}$, due to the coherent interactions among the \an{particles}. 
Thus, we can summarize the ideal phase gate with the mapping 
\cite{QIP:Calarco00b,QIP:Calarco00c}

\begin{eqnarray}
\label{eq:ideal-mapping}
\begin{array}{lccl}
\ket{\epsilon_1}\ket{\epsilon_2}\ket{\psi_{\epsilon_1\epsilon_2}} & 
\rightarrow & e^{i\phi\epsilon_1\epsilon_2} & \!\!\!\ket{\epsilon_1}
\ket{\epsilon_2}\ket{\psi_{\epsilon_1\epsilon_2}},
\end{array}
\end{eqnarray} 
where the motional state $\ket{\psi_{\epsilon_1\epsilon_2}}$ has to factor out 
at the beginning and at the end of the gate operation. In the ideal 
transformation (\ref{eq:ideal-mapping}) we grouped together the kinematic 
and global two-particle phases. Indeed, the application of single-qubit 
operations affords $\phi_g = \phi_{11} - \phi_{01} - \phi_{10} +
\phi_{00}$ \cite{QIP:Calarco01}.  

We conclude this section by introducing the concept of gate fidelity
$F\in[0,1]$, which will be a useful quantity later to assess the gate 
performance. Basically, it is the projection of the physical state 
obtained by actually manipulating the system onto the logical state that the 
gate aims to attain, averaged over degrees of freedom (e.g., motion) that
cannot be accurately controlled.  


\section{Two-qubit gates based on collisional interactions}

In this section we analyze in some detail two important groups of \an{gate schemes with neutral atoms 
that are well suited for atom chip based implementations}: 
the first one encodes the quantum information in internal levels of the atoms, and utilizes the 
external (motional) degrees of freedom to manipulate the information; the second one, instead, stores  
the information in the external degrees of freedom and the internal ones might be used to process it. 
Importantly, in both scenarios, the entanglement between the qubits is produced by the collisions 
among the atoms.

In order to obtain conditional dynamics, as we discussed in the previous section,
either the trapping potential or the interaction term should be state-dependent. 
In the case of ultra-cold neutral atoms, the interaction between atoms is mediated by 
two-body collisions, whose dominant contribution is s-wave scattering 
described by 
\begin{eqnarray}
\label{eq:coll-int}
U_{k\ell}({\bf r},{\bf r}^{\prime}) = 
\frac{4\pi\hbar^2a_{\rm s}^{k\ell}}{m}\delta^3({\bf r} - {\bf r}^{\prime}), 
\end{eqnarray}
where $a_{\rm s}^{k\ell}$ is the s-wave scattering length for the internal states
$\ket{k}$ and $\ket{\ell}$. 
Due to the short range of the pseudopotential (\ref{eq:coll-int}), the wave 
functions of the atoms have to overlap in order to interact, and for identical 
atoms in the same logical state, s-wave scattering is only possible for
bosons, and therefore in the following we will consider bosonic atomic species. 
As the potential given in Eq. (\ref{eq:coll-int}) assumes elastic collisions, the 
states $\ket{0}$ and $\ket{1}$ have to be chosen such that they remain the 
same after the collision.

\subsection{Internal-state qubits}
\label{QIP:sec:intstate_collint}

One of the most effective theoretical models for an atom chip phase gate 
has been proposed in Ref.~\cite{QIP:Calarco00a}. 
In this scheme the control of the interaction between the atoms is realized 
by changing the shape of a microscopic potential depending on the internal 
state of the atoms. Three conditions are assumed: $(i)$ the 
shape of the potential is harmonic; $(ii)$ the atoms are initially cooled to the 
vibrational ground state of two potential wells centered at ${\bf r} = {\bf r}_0$ 
and ${\bf r} = -{\bf r}_0$; $(iii)$ the change in the form of the trapping 
potential is instantaneous. The principle of the gate is the following: at 
time $t=0$ the barrier between the atoms, say in the $x$ direction, 
is suddenly removed (selectively) for atoms in the logical state 
$\ket{1}$, whereas for atoms in the internal state $\ket{0}$ the potential 
is not changed. An atom in the logical state $\ket{1}$ finds itself in a new
harmonic potential centred at ${\bf r} =0$ with a frequency $\omega$, smaller  
than the one of the separated wells, $\omega_0$. The atoms in state $\ket{1}$ 
are allowed to perform an integer number of oscillations and at $t=\tau_g$ the initial wells are
restored. In this process the particles acquire both a kinematic phase due 
to their oscillations in the traps and an interaction phase due to their 
collisions. In the tight transverse confinement regime, where the frequency 
($\omega_{\perp}$) of the well in the $y,z$ directions is much larger than
that ($\omega,\omega_0$) in the $x$ direction, the gate dynamics can be 
well approximated by a one-dimensional (1D) model with a contact potential 
$U_{k\ell}^{\rm 1D}(x,x^{\prime}) = 2\hbar\omega_{\perp}a_{\rm
  s}^{k\ell}\delta(x - x^{\prime})$ \cite{QIP:Petrov00}. The kinematic phase
$2n\pi\omega_{\perp}/\omega$ ($\tau_g = 2\pi n/\omega$, $n\in\mathbb{N}$) due to the radial 
confinement is common to all states, while the one due to the oscillation in 
the axial direction is state dependent. Due to the harmonicity of the trapping 
potentials, almost perfect revivals of the wave packet occur. By choosing $\omega = 
\omega_0/j$, with $j\in\mathbb{N}$, and in the limit where the interaction 
does not induce any relevant alteration in the shape of the two-particle wave 
function, the gate operation time $\tau_g$ can be fixed by looking at the revival 
where the total accumulated phase $\phi_g$ assumes a defined value, e.g., $\pi$. 
Due to the form of $U_{k\ell}^{\rm 1D}$ the frequency $\omega_{\perp}$ can be
adjusted in order to fix the value of $\phi_g$ \cite{QIP:Calarco00a}. 

Atom chips can provide microscopic
state dependent potential landscapes in which atoms can be 
trapped and manipulated for the implementation of the above scheme. 
In Ref.~\cite{QIP:Schmiedmayer02} it was pointed out that a combination 
of static magnetic and static electric fields could be used for this purpose.
However, several issues have to be addressed that could prevent a 
successful experimental realization of the scheme discussed above:
($i$) the trapping potentials are usually anharmonic;
($ii$) the fidelity is strongly reduced by wave packet distortion due 
to undesired collisions in some of the qubit basis states
\cite{QIP:Calarco00a}; ($iii$) transverse excitations of the atoms can 
arise during the collisions if the ratio 
$\omega_{\perp}/\omega$ is not properly chosen for the 1D condition. Those
processes would significantly reduce the gate fidelity. 

An analysis of the limitations due to anharmonicity of the potentials 
is carried out in Ref.~\cite{QIP:Negretti05}. In that analysis a cubic 
and a quartic term is added to the harmonic potential in order to 
include the next leading order terms in the Taylor
series expansion of an arbitrary potential. While a cubic
anharmonicity is well tolerated, the quartic correction poses severe 
restrictions to the correct performance of the gate that are not 
easy to satisfy on atom chips. Thus, for a correct performance, 
the atoms have to be forced to a given dynamics.    

\begin{figure}[tb]
\includegraphics[width=0.48\textwidth]{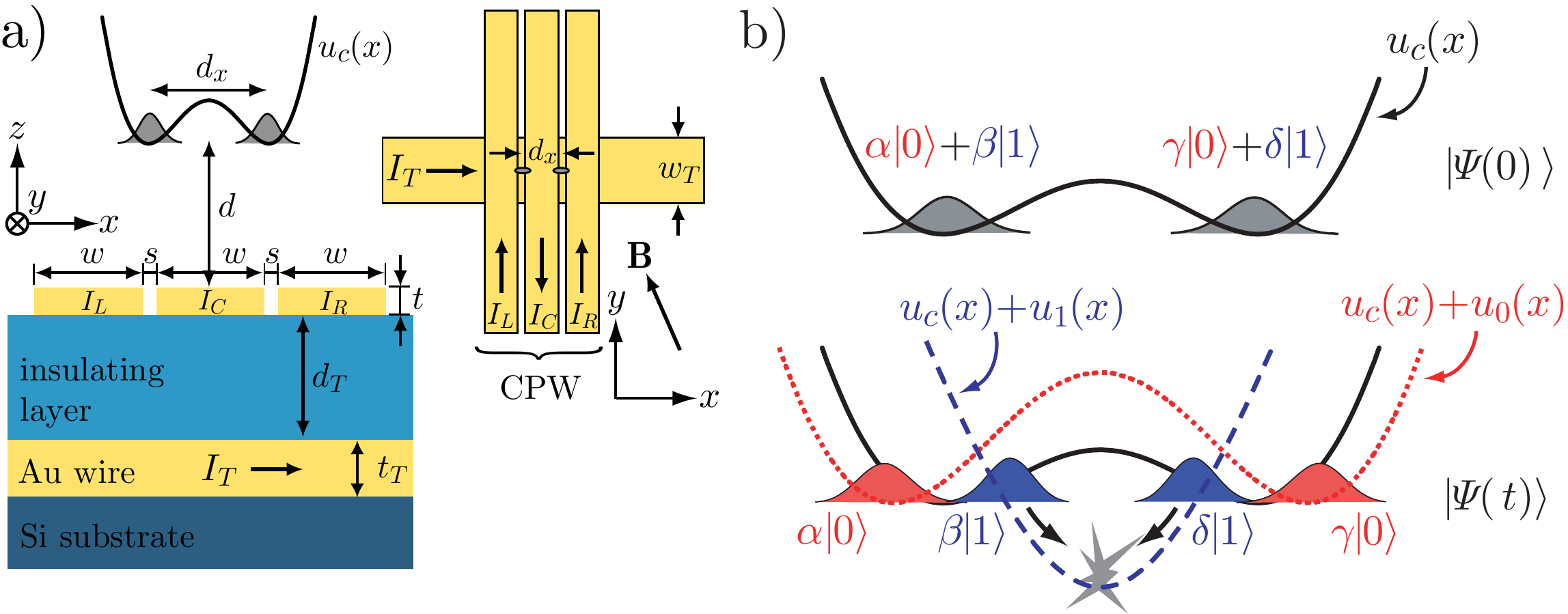}
\caption{a) Layout of the atom chip for the microwave collisional
  phase gate. b) State-selective potential, atomic wave functions, and principle of 
the gate operation. Top: the state-independent potential $u_c(x)$ along $x$.
Bottom: the state-dependent potential $u_c(x) + u_k(x)$ [here
$\lambda(t)=1$]. The 
atomic wave functions after half an oscillation period are shown~\cite{QIP:Treutlein06b}. 
Copyright (2006) by The American Physical Society.} 
\label{QIP:fig:Results}
\end{figure}

The variant of Ref.~\cite{QIP:Treutlein06b} to the original proposal
\cite{QIP:Calarco00a} can be regarded as the first attempt towards a realistic 
implementation of the collisional phase gate on an \an{atom-chip device}. 
It employs the hyperfine qubit states $|0\rangle\equiv\ket{F=1,m_F=-1}$ and 
$|1\rangle\equiv\ket{F=2,m_F=1}$ of $^{87}$Rb whose favorable coherence 
properties were already discussed in Sec.~\ref{QIP:sec:qubitstatesHyper}. 
\tunnu{Moreover, its key ingredient, the coherent manipulation of these states 
with a state-dependent trapping potential and the control of collisions, was realized in recent 
experiments with BECs \cite{QIP:Boehi09,QIP:Riedel10}.} Let us analyze the
features of this scheme (see Fig. 
\ref{QIP:fig:Results}). The state-dependent potential is split into 

\begin{eqnarray}
\label{eq:StateSelPot}
V_k(\mathbf{r},t) = u_c(\mathbf{r}) + \lambda(t) u_k(\mathbf{r}),
\end{eqnarray}
where $u_c(\mathbf{r})$ is a common part and $u_k(\mathbf{r})$ a
qubit-state dependent part ($k =0,1$). The common part 
of the potential is a time-independent double well potential along 
$x$ that can be realized by a static magnetic potential, which is 
nearly identical for the chosen qubit states. As in
Ref.~\cite{QIP:Calarco00a}, the dynamics can be reduced to 1D assuming a tight
confinement in the transverse dimensions $y$, $z$. The
state-dependent part can be realized by a microwave near-field potential (see below). It is modulated with a function 
$\lambda(t)$, with $0 \leq \lambda(t) \leq 1$. At times 
$t<0$, when the gate is in its initial state, we have $\lambda(t) = 0$ and 
the atoms are subject to $u_c(\mathbf{r})$ only.  Each atom is
prepared in the motional ground of one of the wells of the double well potential. During the time 
$0\leq t \leq \tau_g$, $\lambda(t) \neq 0$ and the potential is state-dependent. 
The effect of $u_k(\mathbf{r})$ is twofold: $u_1(\mathbf{r})$ removes the
barrier of the double well for state $|1\rangle$ and atoms in this state start
to oscillate; the potential $u_0(\mathbf{r})$ shifts the minima of the double 
well for state $|0\rangle$ further apart in the $x$-direction [see Fig.~\ref{QIP:fig:Results}b)], whereas in the 
original proposal those atoms do not experience any trap change. In this way, 
unwanted collisions (atoms in state $|01\rangle$), which are a major source 
of infidelity, are strongly reduced and the map (\ref{eq:ideal-mapping}) 
is implemented. 

The state-dependent potential can be realized with microwave near-fields 
with a frequency near the hyperfine splitting of $^{87}$Rb of $6.8$~GHz. 
Unlike the optical potentials created by non-resonant laser beams, 
which can be tightly focussed due to their short wavelength, the centimeter
wavelength $\lambda_\textrm{mw}$ of microwave radiation poses severe
limitations on far-field traps. On atom chips, however, the atoms are trapped at
distances $d\ll \lambda_\textrm{mw}$ from the chip surface, and therefore
they can be manipulated by microwave signals in on-chip transmission lines. 
In the near field of the source currents and voltages, the microwave fields 
have the same position dependence as the static fields created by equivalent 
stationary sources. The maximum field gradients depend on the size of the 
transmission line conductors and on the distance $d$, not on
$\lambda_\textrm{mw}$. \tunnu{Therefore, state-dependent microwave potentials
varying on the micrometer scale can be realized \cite{QIP:Boehi09,QIP:Boehi10}.}  
In a related way, radio frequency fields can be used to generate near-field 
potentials \cite{QIP:Lesanovsky06}.

When we consider the hyperfine levels $|F,m_F\rangle$ of the $5S_{1/2}$ ground 
state of a $^{87}$Rb atom, the magnetic component of the microwave field 
$\mathbf{B}_\mathrm{mw}(\mathbf{r})\cos(\omega t)$ couples the $|1,m_1\rangle$ 
to the $|2,m_2\rangle$ sublevels, with Rabi frequencies
\begin{eqnarray} 
\label{eq:OmegaR}
\Omega_{1,m_1}^{2,m_2} (\mathbf{r}) = \frac{\langle 2,m_2|
  \hat{\boldsymbol{\mu}} \cdot 
\mathbf{B}_\mathrm{mw}(\mathbf{r}) |1,m_1\rangle }{\hbar},
\end{eqnarray}
for the different transitions (in the rotating-wave approximation). 
In Eq. (\ref{eq:OmegaR}), $\hat{\boldsymbol{\mu}}=\mu_B g_J \mathbf{\hat
  J}$ is the operator of the electron magnetic moment ($g_J\simeq 2$). In a 
combined static magnetic and microwave trap, as considered here, both the static field
$\mathbf{B}_s(\mathbf{r})$ and the microwave field $\mathbf{B}_\mathrm{mw}(\mathbf{r})$ vary 
with position. This leads to a position-dependent microwave coupling with 
in general all polarization components present. The detuning of the microwave 
from the resonance of the transition $|1,m_1\rangle \rightarrow |2,m_2\rangle$ is: 

\begin{eqnarray}
\Delta_{1,m_1}^{2,m_2}(\mathbf{r}) = \Delta_0 - \frac{\mu_B}{2\hbar}(m_2 +
m_1) |\mathbf{B}_s(\mathbf{r})|,
\end{eqnarray}
where $\Delta_0 = \omega - \omega_0$ is the detuning from the transition 
$|1,0\rangle \rightarrow |2,0\rangle$, and the different Zeeman 
shifts of the levels have been taken into account. The limit of 
large detuning $|\Delta_{1,m_1}^{2,m_2}|^2 \gg |\Omega_{1,m_1}^{2,m_2}|^2$
allows for long coherence lifetimes of the qubit states in the microwave
potential. In this limit, the magnetic microwave potentials for the sublevels of $F=1$ 
(left) and $F=2$ (right) are given by
\begin{eqnarray}
V_\mathrm{mw}^{1,m_1}(\mathbf{r}) &=& \frac{\hbar}{4}\sum_{m_2}{
  \frac{|\Omega_{1,m_1}^{2,m_2}(\mathbf{r})|^2} 
{\Delta_{1,m_1}^{2,m_2}(\mathbf{r})} },\nonumber\\
V_\mathrm{mw}^{2,m_2}(\mathbf{r}) &=& - \frac{\hbar}{4}\sum_{m_1}{
  \frac{|\Omega_{1,m_1}^{2,m_2}(\mathbf{r})|^2} 
{\Delta_{1,m_1}^{2,m_2}(\mathbf{r})} }.
\end{eqnarray}
As desired, the potentials for $F=1$ and $F=2$ have opposite signs, leading to
a differential potential for the qubit states $|0\rangle\equiv\ket{1,-1}$ and 
$|1\rangle\equiv\ket{2,1}$. 

In addition to the magnetic microwave field, the electric field 
$\mathbf{E}_\mathrm{mw}(\mathbf{r})\cos(\omega t + \varphi)$ also leads to
energy shifts. By averaging over the fast oscillation of the microwave at 
frequency $\omega$, which is much faster than the atomic motion, the electric 
field leads to a time-averaged quadratic Stark shift. 
Hence, the total microwave potential for state $|0\rangle$, $u_0(\mathbf{r})$ 
in (\ref{eq:StateSelPot}), is
\begin{eqnarray}
u_0(\mathbf{r}) = - \frac{\alpha}{4} |\mathbf{E}_\mathrm{mw}(\mathbf{r})|^2
+ \frac{\hbar}{4}\sum_{m_2=-2}^{0}{
  \frac{|\Omega_{1,-1}^{2,m_2}(\mathbf{r})|^2} 
{\Delta_{1,-1}^{2,m_2}(\mathbf{r})} },
\end{eqnarray}
while the microwave potential for state $|1\rangle$ is 
\begin{eqnarray}
u_1(\mathbf{r}) = - \frac{\alpha}{4} |\mathbf{E}_\mathrm{mw}(\mathbf{r})|^2
- \frac{\hbar}{4}\sum_{m_1=0}^{+1}{
  \frac{|\Omega_{1,m_1}^{2,+1}(\mathbf{r})|^2} 
{\Delta_{1,m_1}^{2,+1}(\mathbf{r})} }.
\end{eqnarray}

The atom chip layout shown in Fig.~\ref{QIP:fig:Results}a) allows one to 
realize the desired state-selective potential. It consists of two layers of gold 
metallization on a high resistivity silicon substrate, separated by a thin dielectric 
insulation layer. The wires carry stationary (DC) currents, which, when combined 
with appropriate stationary and homogeneous magnetic bias fields, create the
state-independent potential $u_c(\mathbf{r})$. In addition to carrying DC
currents, the three wires on the upper gold layer form a coplanar waveguide
(CPW) for a microwave at frequency $\omega$, which is possible by the
use of bias injection circuits. The microwave fields guided by
these conductors create the state-dependent potential $u_k(\mathbf{r})$. 

\toni{
In order to speed up the correct gate performance, optimal control techniques 
\cite{QIP:Calarco04} have been employed. To introduce the related concepts, 
let us consider a system governed by the time dependent Hamiltonian $\hat H(t,\lambda)$, 
where $\lambda$ is the control parameter. Our goal in general is to reach, in a fixed time 
$\tau_g$ \an{and for a given initial condition $\ket{\psi_0}$ at time $t=0$}, a 
certain target state $|\psi_{\rm T}\rangle$ with high fidelity. Several iterative 
algorithms exist that can yield a systematic improvement in the gate fidelity. 
Here the one developed by Krotov \cite{QIP:Krotov96,QIP:Sklarz02} has been used. The 
quantum optimal control algorithm works as follows:}

\begin{figure}[tb]
\includegraphics[width=0.48\textwidth]{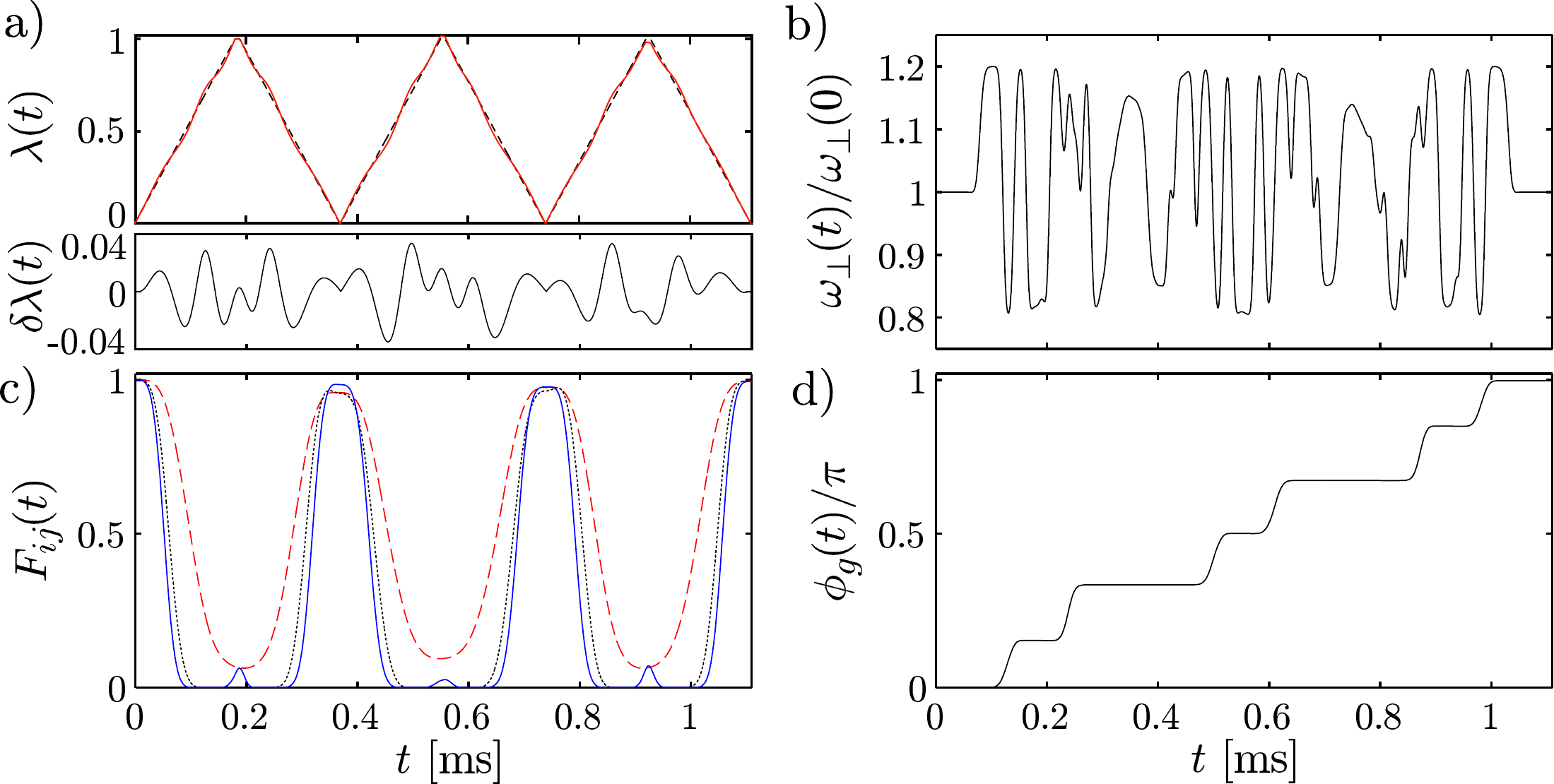}
\caption{\toni{Dynamics during the gate operation, shown for $N=3$ oscillations~\cite{QIP:Treutlein06b}. The gate time is $\tau_g=1.11$ ms. 
Both $\lambda(t)$ and $\omega_{\perp}(t)$ are used as control parameters. a) Optimal control of the microwave 
power. Upper plot: initial trial function $\lambda^{(0)}(t)$ (dashed line) and optimized control parameter $\lambda(t)$ 
(solid line). Each triangular ramp of $\lambda(t)$ corresponds to a full oscillation of state $\ket{1}$. Lower plot: 
the difference $\delta\lambda(t)=\lambda(t)-\lambda^{(0)}(t)$ shows small modulations. b) Optimal control of the 
effective one-dimensional interaction strength via modulation of the transverse trap frequency $\omega_{\perp}$. 
c) Evolution of the overlap fidelities during the gate operation: $F_{00}(t)$ (dashed line), $F_{01}(t)=F_{10}(t)$ 
(dotted line), and $F_{11}(t)$ (solid line), where $F_{ij}(t) = \vert\langle \psi_{ij}(x_1,x_2,t)|\psi(x_1,x_2,0)\rangle\vert^2$ 
with $ \psi_{ij}(x_1,x_2,t)$ being the two particle state. d) Evolution of the gate phase $\phi_g(t)$. The phase shift 
steps are due to the six collisions in state $\ket{11}$. Copyright (2006) by The American Physical Society.}} 
\label{QIP:fig:optimized}
\end{figure}

\begin{enumerate}
\item \toni{an initial guess $\lambda^{(0)}(t)$ is chosen for the control
parameter. 
\item the initial state $|\psi_0\rangle$ is evolved in
time according to the Schr\"odinger equation
$|\psi_{\lambda}(\tau_g)\rangle=\hat{\mathcal{U}}(\lambda,\tau_g)|\psi_0\rangle$ until
time $\tau_g$. 
\item an auxiliary state
$|\chi_{\lambda}(\tau_g)\rangle\equiv|\psi_{\rm T}\rangle\langle\psi_{\rm
T}|\psi_{\lambda}(\tau_g)\rangle$ is defined, which can be interpreted as
the part of $|\psi_{\lambda}(\tau_g)\rangle$ that has reached the objective
$|\psi_{\rm T}\rangle$; the auxiliary state is evolved backwards in time until $t=0$.
\item $|\chi_{\lambda}(t)\rangle$ and $|\psi_{\lambda}(t)\rangle$ are propagated 
again forward in time, while the control parameter is updated $\lambda^{(j+1)}(t)=
\lambda^{(j)}(t)+2/\eta(t)\cdot\Im\left[\langle\chi^{(j)}_{\lambda}(t)|\partial_{\lambda}
\hat H|\psi^{(j+1)}_{\lambda}(t)\rangle\right]$, where $j$ refers to the $j$-th iteration of 
the algortihm. The weight function $\eta(t)$ constrains the initial and final values of 
the control parameter.
\item steps 3. and 4. are repeated until the desired value of the fidelity is obtained.}
\end{enumerate}\toni{
Making use of this technique, a gate operation time $\tau_g = 1.11$~ms with a 
fidelity $F = 0.996$ can be obtained (see also Fig. \ref{QIP:fig:optimized}), as 
shown in Ref.~\cite{QIP:Treutlein06b}, \an{the control parameters being the 
electrical microwave current injected in the CPW and the radial frequency 
$\omega_{\perp}$ through the stationary electrical currents in the wires 
displayed in Fig.~\ref{QIP:fig:Results} a).} We emphasize that with this 
$\tau_g$ and the long coherence lifetime of the qubit pair chosen, 
thousands of gate operations can be accomplished. 
The fidelity calculation includes the effect of several error sources: trap 
losses and decoherence due to the chip surface, undesired two-photon
transitions induced by the microwave, mixing of the 
hyperfine levels due to the microwave coupling, and qubit dephasing due to technical noise. In the
limit of large microwave detuning, the admixture of 
other states with different magnetic moments to the qubit states is strongly reduced. A last important point is
related to the difficulty to prepare the atoms in the vibrational ground 
state with close to 100\% efficiency. This effect, modeled by a finite temperature, 
has been also included in the analysis. 
For temperatures $T \leq 20$~nK in the
initial double well trap, the fidelity is not reduced significantly \cite{QIP:Treutlein06b}.}

\subsection{Motional-state qubits}
\label{sec:motional}

Based again on the conditional phase shifts induced by the collision between
cold atoms, a number of proposals rely on the manipulation of quantum
information stored in motional degrees of freedom. The original proposals 
\cite{QIP:Eckert02,QIP:Charron02,QIP:Mompart03} dealt with optical lattices, 
but those schemes can also be implemented using microscopic potentials 
on chips. 
Besides magnetic, microwave, or radio frequency traps, chip-based optical traps  are of interest in this context. 
By illuminating a 2D array of refractive or diffractive micro lenses with laser light, a 2D 
set of diffraction limited laser foci can be formed. Atoms can be confined 
in the optical dipole potentials generated by the laser foci 
\cite{QIP:Birkl07}. In a first experiment, arrays with more than 80 sites 
were loaded with ensembles of about 1000 trapped $^{85}$Rb atoms in the 
centre of the 2D configuration and about 100 at the edges \cite{QIP:Dumke02}. 
Theoretical proposals for two-qubit gates in this system rely again upon the 
spatial overlap of two qubits out of initially separated 
locations. This can be accomplished by illuminating the array of micro lenses 
with two laser beams with a finite relative angle of propagation creating 
two interleaved sets of dipole traps \cite{QIP:Birkl07}. \ann{The variation of 
the relative angle yields a variation in the mutual distance between the 
trap sets or, alternatively, by using a programmable and computer 
controllable nematic liquid-crystal spatial light modulator, the trap 
separation can be varied by changing the periodicity of the modulator~\cite{QIP:Bergamini04}.} 
An important feature of these optical micropotentials
is the relatively large separation of neighbouring sites ($\sim 125~\mu$m) 
which enables individual addressing \cite{QIP:Dumke02}. 
In a recent experiment the same group has demonstrated that transport, 
reloading, and a full shift register cycle ($\sim 55\mu\mathrm{m}$) can 
be performed with negligible atom loss, heating, or additional dephasing 
or decoherence \cite{QIP:Lengwenus10}. This proves that such a technology 
is scalable to complex and versatile 2D architectures such that quantum 
information processing, quantum simulation, and multi-particle entanglement 
become accessible. 

In the quantum gate scheme of Ref.~\cite{QIP:Charron02}, a double 
well potential contains one atom per well. The logic states 
$\ket{0}$ and $\ket{1}$ are identified with the single particle ground and 
excited states of each well, respectively. Initially the barrier is
sufficiently high that tunneling between the 
lowest four eigenstates of a single trapped atom is negligible. When the 
barrier is lowered in such a way that the single particle excited states 
(the qubit state $\ket{1}$) of the potential do overlap, tunneling takes place 
and the energy shift due to the atom-atom interaction increases exponentially. 
The interaction lasts for a time sufficient to 
accumulate the required phase shift for a phase gate and subsequently the initial 
trapping configuration is restored by increasing the barrier again. An accurate 
use of quantum interference between two-particle states yields an optimised 
gate duration of 38~ms with an infidelity $1-F \approx 6.3\times
10^{-6}$. In the proposal of Ref.~\cite{QIP:Eckert02} the scenario is very
similar and it uses the same qubit set. While in Ref.~\cite{QIP:Charron02} 
the two-qubit gate is physically realised by lowering and increasing the 
barrier of the double well potential, in Ref.~\cite{QIP:Eckert02} the (initially) 
separated traps adiabatically approach (or separate from) each other. 
In that way it is possible to obtain $\tau_g \sim 20$ ms for a $\sqrt{{\rm SWAP}}$ 
two-qubit gate. 

The proposal of Refs.~\cite{QIP:Eckert02,QIP:Mompart03} uses again motional 
states, but not strictly the vibrational states of the trap. In the scheme 
each qubit consist of two separated traps and a single atom. Now, the 
computational basis $\{\ket{0},\ket{1}\}$ is formed in this way: the ground 
state of the left trap represents $\ket{0}\equiv\ket{0}_{\rm L}$, whereas the ground 
state of the right trap represents $\ket{1}\equiv\ket{0}_{\rm R}$. One- 
and two-qubit quantum gates are performed by adiabatically approaching the 
trapping potentials and allowing for tunneling to take place. We note that in 
such a scheme four wells are needed to implement a two-qubit gate, either 
arranged in a 1D configuration with the traps on a line or side-by-side 
in a 2D configuration. Taking into account the different error sources 
present in this scheme, like fluctuations of the trap positions, photon 
scattering, and heating, one obtains an error rate of about 0.02, with a 
single-qubit operation time of 4 ms, and $\tau_g \sim 10$ ms for a two-qubit 
operation such as a phase gate. Even though the error rate is rather large, 
the scheme offers several advantages: 1) decoherence due to spontaneous 
emission reduces the fidelity only marginally; 2) no momentum transfer is 
effected for single and two-qubit gates; 3) a state-dependent interaction 
is not required for the implementation of two-qubit gates; 4) the readout 
is done with a laser beam focused onto one trap minimum and detecting 
the fluorescence light; 5) since one- and two-qubit gates are realized using 
the same technique, i.e., by approaching the traps adiabatically, the complexity 
of the experimental setup would be reduced. 

Both of these schemes can be implemented on a chip by means of a
combination of static and microwave fields, with the need of trapping only 
one hyperfine level. However, one can combine the nice coherence properties 
of the qubit states $\ket{F=2, m_F=1}$ and $\ket{F=1, m_F=-1}$ of the ground 
state of $^{87}$Rb, and the entanglement produced by cold 
collisions via the motional states ($\ket{g},\ket{e}$). In Ref.~\cite{QIP:Cirone05} 
two different ways of realizing this concept have been proposed: (a) 
duplicate the logical state of the storage levels in the motional levels, 
where $\ket{1g}\leftrightarrow\ket{1e}$; (b) swap the logical states of 
the two degrees of freedom, $\ket{g1} \leftrightarrow\ket{e0}$. 
Here we consider only the swap scheme. 

\begin{figure}[tb]
\includegraphics[width=0.48\textwidth]{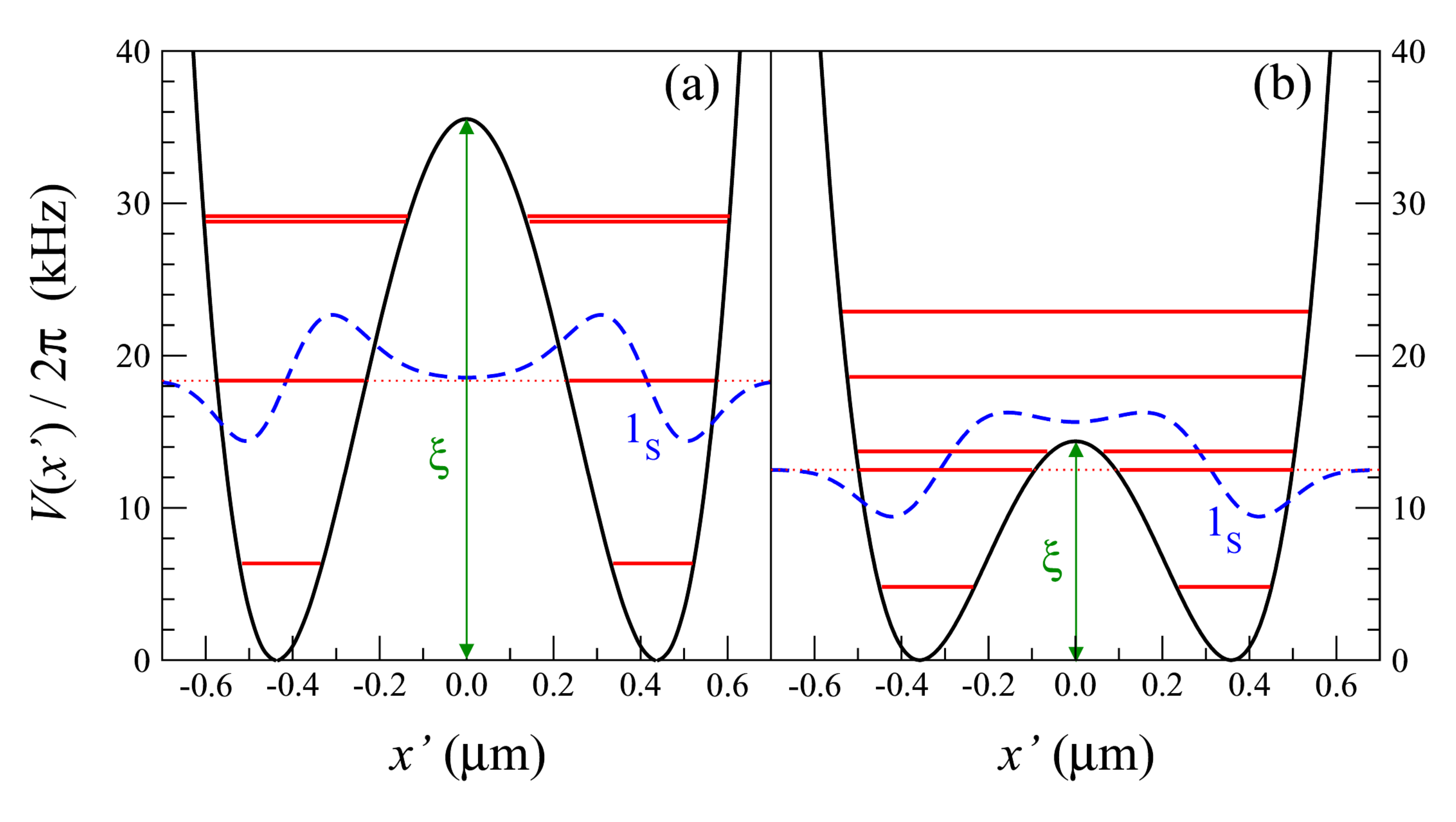}
\caption{\toni{Double-well potentials created by a realistic atom-chip configuration. 
The energies of the first six eigenstates 
are shown as red (horizontal) lines. The blue (dashed) line represents the wave 
function of the third eigenstate labeled as $1_S$ because it originates from the 
symmetric combination of the first excited trapped levels, also labeled as $\ket{e}$ 
in the text. a) Highest barrier ($\xi$); b) lowest barrier (see Ref.~\cite{QIP:Charron06} for details). 
Copyright (2006) by The American Physical Society.}} 
\label{QIP:fig:dwp}
\end{figure}

Given the initial state 

\begin{equation}
\ket{\varphi_0} = \left( a\ket{00} + b\ket{01}
+ c\ket{10} + d\ket{11} \right) \ket{gg}, 
\end{equation}
the swap scheme takes place in three steps: 

\begin{enumerate}
\item we selectively excite the operation state and 
de-excite the storage states $\ket{\varphi_1'} =\ket{00} \left( a\ket{gg} + b\ket{ge}
+ c\ket{eg} + d\ket{ee} \right)$, i.e, we swap their logic states; 
\item the operation states get a dynamical phase $\phi_g$:
$\ket{\varphi_2'} =\ket{00} \left( a\ket{gg} + b\ket{ge}
+ c\ket{eg} + e^{i\phi_g} d\ket{ee} \right)$ through collisions; 
\item we swap again the storage and operation states $\ket{\varphi_3'} =\left( a\ket{00} + b\ket{01}
+ c\ket{10} + e^{i\phi_g} d\ket{11} \right) \ket{gg}$.
\end{enumerate}
Such a swap gate scheme is not restricted to internal 
and external degrees of freedom of cold atoms, but it can be applied to any
system with at least two degrees of freedom. A selective excitation
of vibrational states is required when cold atoms are employed. In order to realize
it, the use of two-photon Raman transitions has been suggested. The
experimental implementation of such transitions with $^{87}$Rb atoms is 
rather delicate and a careful analysis is given in Ref.~\cite{QIP:Charron06}. 
In a static version of the swap scheme, where the barrier is fixed and it is 
designed in such a way that the left and right single particle excited states 
overlap (see also Fig. \ref{QIP:fig:dwp}), an operation time for a phase gate of 16.25 ms with a gate fidelity 
$F>0.99$ has been predicted \cite{QIP:Cirone05}. In an optimized version of the 
gate dynamics, where the barrier is lowered and increased in order to get 
faster operation times for a desired value of infidelity, it has been possible 
to achieve fidelities of 0.99 in 6.3~ms and of 0.999 in 10.3~ms 
\cite{QIP:Charron06}.  


\section{Optical lattice based schemes}
\label{sec:optical}
\an{
In this section we want to discuss the use of optical lattices for quantum computing, 
how the lattice potentials can be moved in a state-selective way for implementing the 
two-qubit gate of Refs.~\cite{QIP:Jaksch99} and how these potentials can be realized on a chip. 
Beside this, we shall describe other QIP implementations with optical lattices.}

\subsection{\an{Hamiltonian for a degenerate quantum Bose gas in an optical lattice}}
\an{
We assume a Bose-Einstein condensate of atoms in the internal state $\ket{0}$ to be
loaded into a potential $V_{\mathrm{opt}}(\mathbf{r})+
V_{\mathrm{ext}}(\mathbf{r})$, where $V_{\mathrm{opt}}(\mathbf{r}) = \sum_{j=x,y,z}V_j\sin^2(\kappa j)$ is a 
periodic optical lattice potential and $V_{\mathrm{ext}}(\mathbf{r})$ is an additional external 
potential slowly varying in space compared to $V_{\mathrm{opt}}(\mathbf{r})$. Here 
$\kappa$ is the wave number of the lasers producing the lattice potential. The many-body 
Hamiltonian in second quantization reads as Eq.~(\ref{eq:many-bodyH}), but the sums reduce 
only to the $k=0$ contribution, with the substitution $V_k\rightarrow V_{\mathrm{opt}}+V_{\mathrm{ext}} - \mu$, 
where $\mu$ is the chemical potential. Expanding the field operators in the Wannier basis (while keeping only 
the dominant terms), Eq.~(\ref{eq:many-bodyH}) reduces to the Bose-Hubbard Hamiltonian \cite{QIP:Jaksch98}
}

\an{
\begin{eqnarray}
\hat H=-J\sum_j\hat b_j^{\dag}\hat b_{j+1}+\sum_j(E_j-\mu)\hat n_j +\frac{U}{2}\sum_j\hat n_j (\hat n_j - 1),
\nonumber\\
\end{eqnarray}
where $\hat n_j = \hat b_j^\dag \hat b_j$ counts the number of bosonic atoms at lattice site $j$ 
($[\hat b_j,\hat b^\dag_k] = \delta_{j,k}$), $J$ is the tunnelling matrix element, $U$ describes 
the (repulsive) interaction between particles at the same lattice site, and 
$E_j\equiv V_{\mathrm{ext}}(\mathbf{r}_j)$ is the value of the slowly varying superlattice potential at site $j$. 
The ratio $U/J$ is controlled by the depth of the optical lattice potential $V_j$. Increasing $V_j$ (via the intensity 
of the trapping lasers) reduces the tunneling matrix element $J$ and increases the repulsive interaction between 
the atoms $U$ \cite{QIP:Jaksch98}.
}

\an{
In order to perform gate operations in optical lattices we have to be able to
selectively fill each lattice site with exactly one particle. This can be achieved 
by making use of the phase transition from a superfluid BEC to a Mott 
insulator (MI) at low temperatures (experimentally demonstrated in Ref.~\cite{QIP:Greiner02}), 
which can be induced by increasing the ratio of the onsite interaction $U$ to the tunneling 
matrix element $J$ predicted by the Bose-Hubbard model \cite{QIP:Fisher89,QIP:Bruder93}. 
In the MI phase the density $\rho_j$ (occupation number per site) is pinned at an integer 
$n\in\mathbb{N}$ (when starting from a commensurate filling of the lattice), and thus 
represents an optical crystal with diagonal long range order with period imposed 
by the laser light. Particle number fluctuations are thereby drastically reduced and 
thus the number of particles per lattice site is fixed. The number of particles per lattice 
site depends on the chemical potential in the isotropic case $E_j=0$ \cite{QIP:Fisher89}, 
whereas in the non-isotropic case we may view $\mu-E_j$ as a local chemical potential. Hence, 
$\rho_j$ can be controlled by the external potential. Indeed, the repulsive interaction 
between the particles increases as the optical potential is made deeper. At the same time 
the hopping rate at which particles move from one site to the next decreases. If the optical 
lattice is turned on with a time scale much slower than the hopping rate and if the thermal 
energy $k_BT$ ($k_B$ being the Boltzmann constant) can be kept much smaller than the 
interaction energy between two particles in one site, one can achieve a filling of the optical 
lattice with exactly one particle per lattice site \cite{QIP:Jaksch98}. 
}

\an{
Finally, we note that, recently, by means of quantum optimal control, a significant reduction of the 
preparation time of a MI state with high fidelity has been demonstrated \cite{QIP:Doria11} (up to 
hundreds times shorter than in current experiments). Alternatively, by using the dependence of 
the interaction energy on the vibrational states occupied by the atoms, the entropy of the MI state 
can be also drastically reduced as well as the time for the state preparation \cite{QIP:Sherson11}. 
}

\subsection{\an{State selective moving potentials}}
\begin{figure}[tb]
\includegraphics[width=0.48\textwidth]{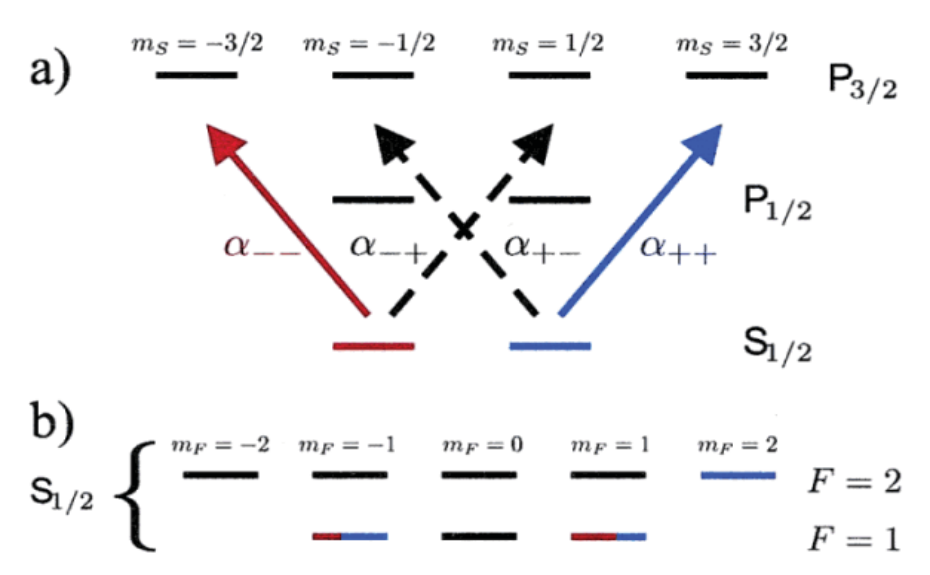}
\caption{\an{a) Fine structure energy levels and laser configuration. The detuning is chosen such 
that the polarizabilities $\alpha_{+-}$ and $\alpha_{-+}$ vanish. b) Hyperfine level structure. Both 
level schemes apply to $^{87}$Rb and $^{23}$Na atoms~\cite{QIP:Briegel00}.}}
\label{QIP:fig:sixs}
\end{figure}

\begin{figure}[tb]
\includegraphics[width=0.48\textwidth]{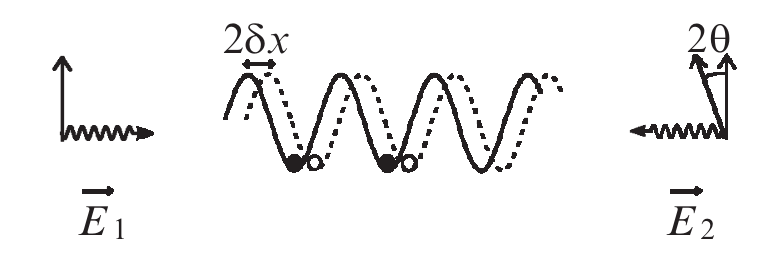}
\caption{\an{Laser configuration along the $z$ axis~\cite{QIP:Briegel00}.}} 
\label{QIP:fig:seven}
\end{figure}

\an{
Following Ref.~\cite{QIP:Jaksch99}, let us consider alkali atoms with a nuclear spin equal to $3/2$ ($^{87}$Rb, $^{23}$Na) 
trapped by standing waves in three dimensions and thus confined by the potential 
$V_{\mathrm{opt}}$. The internal states of interest are hyperfine levels corresponding 
to the ground state $S_{1/2}$ as shown in Fig.~\ref{QIP:fig:sixs}b. Along the $z$ axis, the standing 
waves are in a configuration such that two linearly polarized counter-propagating 
traveling waves with the electric fields $\mathcal{E}_{1,2}$ form an angle 2$\theta$ \cite{QIP:Finkelstein92}, 
as shown in Fig.~\ref{QIP:fig:seven}. 
}

\an{
The total electric field is a superposition of right and left circularly polarized standing waves 
($\sigma^{\pm}$) which can be shifted with respect to each other by changing $\theta$,
}

\an{
\begin{eqnarray}
\!\!\!\!\!\boldsymbol{\mathcal{E}}(z,t)=\mathcal{E}_0e^{-i\nu t}\left[
\boldsymbol{\epsilon}_+\sin(\kappa z +\theta)+
\boldsymbol{\epsilon}_-\sin(\kappa z -\theta)
\right],
\end{eqnarray}
where $\boldsymbol{\epsilon}_{\pm}$ denote unit right and left circular polarization vectors, 
$\kappa=\nu/c$ is the laser wave vector and $\mathcal{E}_0$ the amplitude. The lasers are 
tuned between the $P_{1/2}$ and $P_{3/2}$ levels so that the dynamical polarizabilities $\alpha_{\pm \mp}$ 
of the two fine structure $S_{1/2}$ states, corresponding to $m_s=\pm 1/2$ due to the laser polarization $\sigma^{\mp}$, 
vanish, whereas  the dynamical polarizabilities $\alpha_{\pm \pm}$ due to the laser polarization $\sigma^{\pm}$ are identical 
[see Fig.~\ref{QIP:fig:sixs} a)]. 
Such a configuration can be achieved by tuning the lasers between the $P_{3/2}$ and $P_{1/2}$ fine state levels so that 
the ac-Stark shifts of these two levels cancel each other. The optical potentials for these two states are 
$V_{m_s=\pm 1/2}(z,\theta) = \alpha\vert \mathcal{E}_0\vert^2 \sin^2(\kappa z \pm\theta)$. For instance, if 
$\ket{0}\equiv\ket{F=1,m_F=1}$ and $\ket{1}\equiv\ket{F=2,m_F=2}$, then the potentials for these hyperfine levels are
}

\an{
\begin{eqnarray}
V_{0}(z,\theta)&=&\frac{1}{4}\left[V_{m_s=1/2}(z,\theta)+3V_{m_s=-1/2}(z,\theta)\right],\nonumber\\
V_{1}(z,\theta)&=&V_{m_s=1/2}(z,\theta).
\end{eqnarray}
By using such potentials the atoms can be moved along the $z$ axis in a state-dependent manner and 
the gate scheme proposed in Ref.~\cite{QIP:Jaksch99} can be implemented. In this proposal, at time $t=0$ 
the atom $\alpha$ in the logical state $\ket{k}$ experiences the potential \ann{$V_{\alpha}^{(k)}({\bf r},t) = V(\bar{\bf r}^{(k)}
+\delta {\bf r}_{\alpha}^{(k)}(t) - {\bf  r})$, which is initially ($t<0$) centred at position 
$\bar{\bf r}^{(k)}$. The centres of the potentials move according to the trajectories 
$\delta {\bf r}_{\alpha}^{(k)}(t)$ with the condition 
$\delta {\bf r}_{\alpha}^{(k)}(0) = \delta {\bf r}_{\alpha}^{(k)}(\tau_g) = 0$}, and 
such that the first atom collides with the second one if and only if they are 
in the logical states $\ket{0}$ and $\ket{1}$. 
Such state-dependent potentials can be also realized with 
magnetic microtrap lattices \cite{QIP:Singh08,QIP:Whitlock09}, and with state-dependent 
microwave potentials. The latter can be achieved by using a microwave state dressing scheme where only 
state $\ket{0}$ effectively couples to the microwaves, as e.g. in the experiment of 
Ref.~\cite{QIP:Boehi09}. }

\an{
However, optical lattices can be also realized on a chip. Indeed,} the results of Ref.~\cite{QIP:Christandl04} show
that it is possible to realize 1D and 2D optical lattices, where the
traps are the nodes of the evanescent wave field above an optical waveguide
resulting from the interference of different waveguide modes. With a laser
power of $\sim$ 1 mW it is possible to produce tight traps, 150 nm above the
on-chip waveguide surface, with trap frequencies on the order of 1 MHz, and
with a spatial periodicity of about 1~$\mu$m. Moreover, the individual
qubits are readily addressable, and it is possible to move 1D arrays of qubits
by adjusting the phases of the waveguide modes. The drawback of such
technology is that to get strong confinement with waveguides made from
existing materials and using low laser powers, one needs to work extremely
close to the waveguide surface implying a relevant impact on the qubit
coherence. Alternatively one can use current-carrying wires and a
perpendicularly magnetized grooved structure \cite{QIP:Singh08}. This solution 
allows for the trapping and cooling of ultracold atoms by means of the
current-carrying wires, whereas the magnetic microstructure generates a 1D 
permanent magnetic lattice with a spacing between neighboring sites on the 
order of 10 $\mu$m and with trap frequencies of up to 90 kHz.   

\subsection{Marker qubits}

\begin{figure}[tb]
\includegraphics[width=0.48\textwidth]{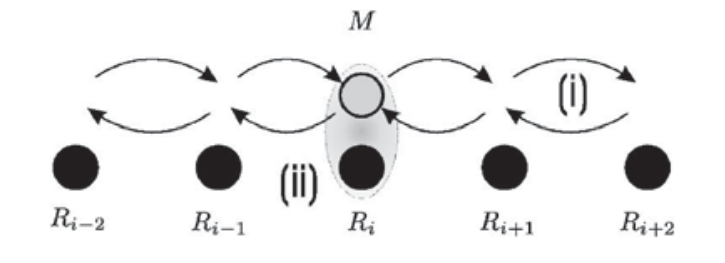}
\caption{\gan{Basic operations with a marker atom on a set of quantum bits (quantum register). 
It can be transported back and forth (i) and it can interact locally with qubits of the quantum register (ii) 
 \cite{QIP:Calarco04}. Copyright (2004) by The American Physical Society.}} 
\label{QIP:fig:markerop}
\end{figure}

\begin{figure}[tb]
\includegraphics[width=0.48\textwidth]{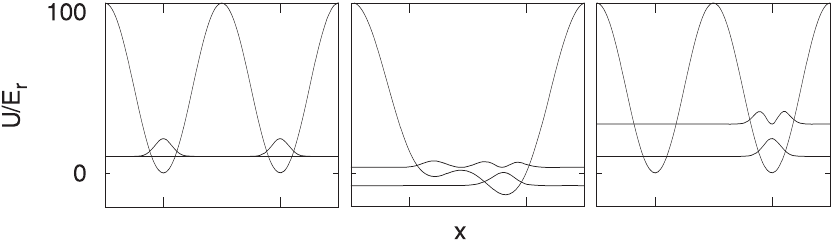}
\caption{\gan{Description of the transport process showing also the atomic wave functions. Left: Initial 
configuration with one atom per lattice site, with the marker 
atom in the left well and the register one in the right well. Center: Intermediate configuration during 
the transport. Right: Final configuration, where the marker atom has been moved in the right well, but in 
a different motional state with respect to the register qubit that remains in the ground state of the well 
\cite{QIP:Calarco04,QIP:Calarco07}. Copyright (2007) by World Scientific.}} 
\label{QIP:fig:marker}
\end{figure}

Another interesting solution to the issue of single-atom addressability via a
laser is quantum computation with neutral atoms, based on the concept of `marker' atoms,
i.e., auxiliary atoms that can be efficiently transported in state independent 
periodic external traps to operate quantum gates between physically distant 
qubits \cite{QIP:Calarco04} \gan{(see Fig.~\ref{QIP:fig:markerop})}. 
Here, again, qubits are represented by internal 
long-lived atomic states, and qubit atoms are stored in a regular array of
microtraps. These qubit atoms remain frozen at their positions during the 
quantum computation. In addition to the atoms representing the qubits, an 
auxiliary `marker atom' (or a set of marker atoms) is considered, 
which can be moved between the different lattice sites
containing the qubits. The marker atoms can either be of a
different atomic species or of the same type as the qubit
atoms, but possibly employing different internal states. These
movable atoms serve two purposes. First, they allow addressing
of atomic qubits by `marking' a single lattice site
due to the marker qubit interactions: the corresponding molecular
complex can be manipulated with a laser without the requirement
of focusing on a particular site. Second, the movable
atoms play the role of `messenger' qubits which allow to
transport quantum information between different sites in the
optical lattice, and thus to entangle distant atomic qubits. 
\gan{The transport process of a marker atom, initially trapped in the left well, 
to the right well occupied by a register qubit is illustrated in Fig.~\ref{QIP:fig:marker}.}
The required qubit manipulation and trapping techniques for such a scheme are essentially 
the same as the ones previously presented, and \an{it is also well suited for chip 
implementations}.

\subsection{\an{Optimal atomic transport in optical potentials}}

\an{
In the above outlined quantum gate schemes, based on the fact that the atomic wave functions 
must be made to overlap, a crucial element is their transport from one site of the optical potential 
to another one, where the interaction among two atoms takes place. This goal has 
to be achieved with a very high efficiency. In Ref.~\cite{QIP:Dechiara08} it has been showed 
that the optimized control sequences allow transport faster and with significantly larger fidelity 
than with processes based on adiabatic transport. This shows again, as we discussed in 
Sec.~\ref{QIP:sec:intstate_collint}, the great potential afforded by the toolbox of quantum optimal 
control for the achievement of high-fidelity quantum computation.}

\an{In the experiment described in Ref.~\cite{QIP:Dechiara08}, a 3D optical lattice has 
been generated by superimposing, say in the $z$ direction, a 1D optical potential to a 2D 
independent optical lattice in the $x-y$ plane. This horizontal lattice can be dynamically
transformed between single-well and double-well configurations, depending on three 
controllable parameters: (i) the depth of the potential wells; (ii) the ratio of vertical to horizontal 
electric field components; (iii) the phase shift ($\theta_b$) between vertical and horizontal light 
components. 
}

\begin{figure}[tb]
\includegraphics[width=0.48\textwidth]{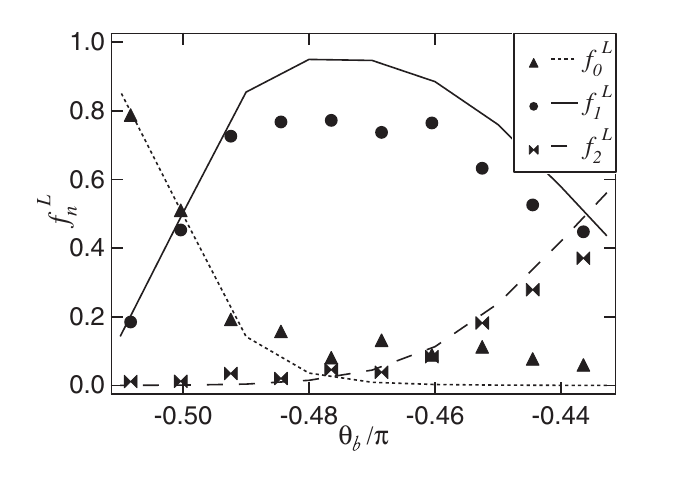}
\caption{\an{Population of the first three
eigenstates of the optical potential at the end of a sequence for shifting the 
atoms from a double- to a single-well configuration as a function of the 
phase shift $\theta_b$ between vertical and horizontal light components~\cite{QIP:Dechiara08} . 
The duration of the sequence is fixed to $T$=0.5 ms. The experimental 
data (symbols) are in good agreement with the numerical data (lines). 
Copyright (2008) by The American Physical Society.}} 
\label{QIP:fig:otto}
\end{figure}

\an{
The assumed theoretical 1D model (in the axial direction) is such that the optical potential can 
be separated along the three spatial directions. This allows to express the atomic wave functions 
as a product of three independent terms. Additionally, like for the collisional gate dimensionality 
conditions discussed in Sec.~\ref{QIP:sec:intstate_collint}, it has been assumed that in the radial 
confinement the atoms always occupy the lowest vibrational state. In Fig.~\ref{QIP:fig:otto} the overlaps 
$f_n^{\alpha}\equiv p_n(T)=\vert\langle\phi_n(T)\vert\hat U(T)\vert\psi_{\alpha}\rangle\vert^2$ of the energy eigenstates 
$\phi_n$ of the final potential with the evolved state $\psi_{\alpha}$ are displayed. Here $\alpha = L , R$ 
indicates the initial well occupancy of the double-well potential, whereas $\hat U(T)$ is the single-particle 
time-evolution operator from time $t = 0$ to the final time $T$. Importantly, the experimental data and 
the theoretical model are in satisfactory agreement, which proves the accuracy of the theoretical modeling. 
Beside this, both the theory and the experiment show a strong dependence on $\theta_b$ for the transport
 of the atom starting in the left site of the double well.
}

\begin{figure*}[htb]
\includegraphics[width=170mm]{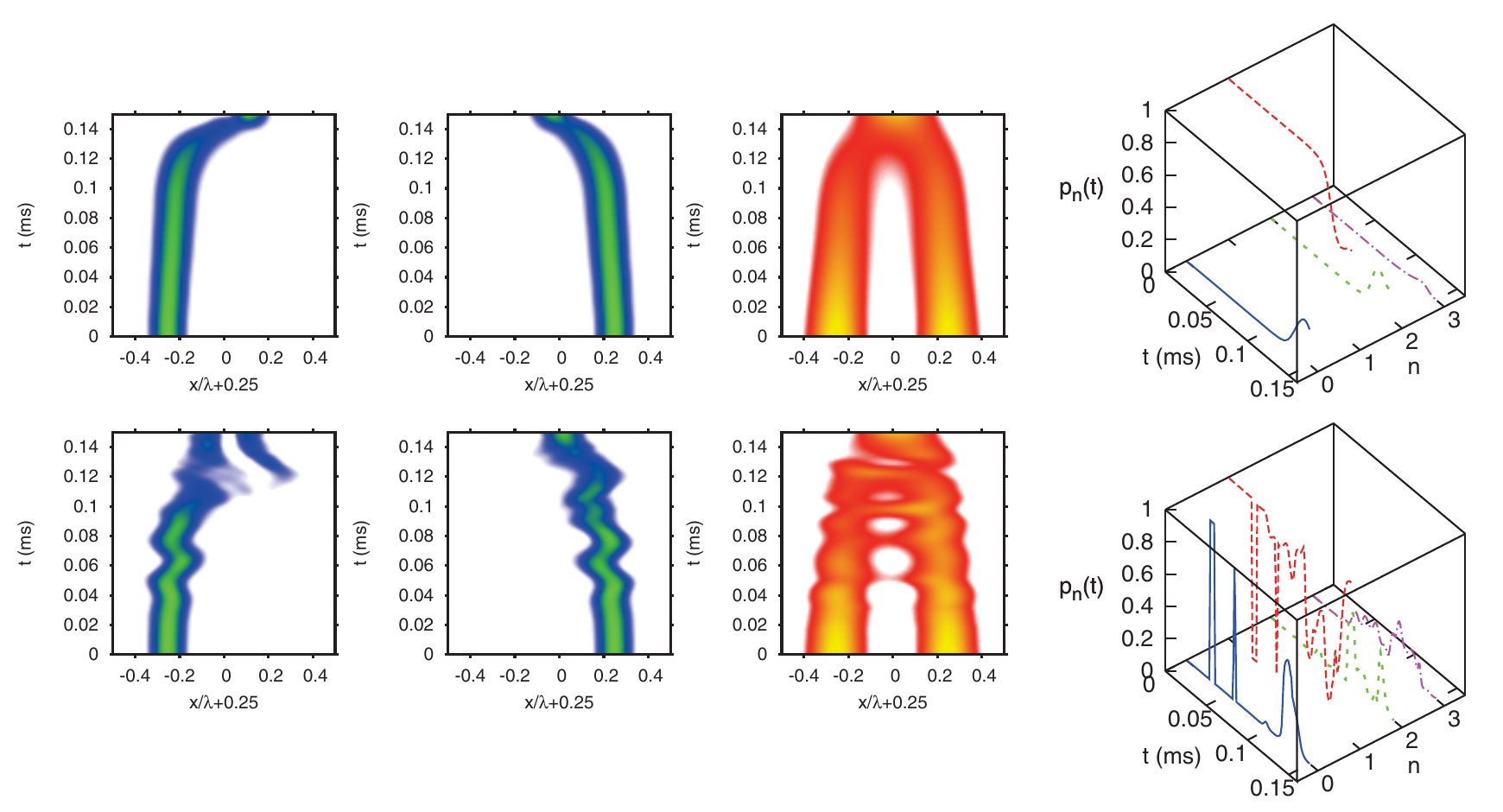}
\caption{\gan{
Comparison between the evolution of the atoms with and without optimal control in the process 
of the transport of two atoms occupying adjacent wells into the same well, a crucial ingredient for the 
realization of a two-qubit quantum gate in the scheme of Ref.~\cite{QIP:Dechiara08}. Top (left to right): nonoptimized case, 
absolute square value of the wave functions as a function of time (atoms initially in the left and right 
well respectively); 1D trapping potential as a function of time; projection $p_n(t)$ of the state initially 
in the left well onto the instantaneous eigenstates $\ket{\phi_n(t)}$ of the confining potential with $n=0$ 
(blue solid), $n=1$ (red dashed), $n=2$ (green dotted), and $n=3$ (magenta dot-dashed). Bottom: 
analogous plots for the optimized case. Copyright (2008) by The American Physical Society.
}} 
\label{QIP:fig:nove}
\end{figure*}

\an{
On the other hand, in Fig.~\ref{QIP:fig:nove} the evolution of the single particle probability density is shown. 
As the figure shows, the optimal time evolution is much less smooth than the adiabatic one, since it 
takes advantage of quantum interference between nonadiabatic excitation paths to obtain better results. 
Furthermore, an analysis of the effect of atom-atom interactions on the transport process shows that the 
optimal control parameter sequences found in the noninteracting case, as the one of Fig.~\ref{QIP:fig:nove}, still 
work when including interaction. Indeed, it has been possible to obtain the same transformation as in the case of 
the adiabatic transport with a better fidelity and in a time shorter by more than a factor of 3, which represents 
a relevant improvement in terms of scalability of the number of gates that can be performed before the system 
decoheres due to the coupling to its environment.}

 \lmn{We also mention, that very recently other schemes for moving atoms in optical lattices has been 
 proposed and that the topic has been reinvigorated by the recent experimental achievements on the 
 realization of single qubit rotations~\cite{QIP:Lundblad09}, on the single site addressability~\cite{Bakr2010,Weitenberg2011}, 
 and on the single atom detection~\cite{QIP:Sherson10,QIP:Bakr09}, as we also discussed in Sec.~\ref{QIP:sec:qubitrothyer}. 
 For instance, one can use optical tweezers, as discussed in Sec.~\ref{sec:motional}, to transport 
 atomic qubits around the lattice and merge them into single lattice sites in order to implement collisional
quantum gates~\cite{QIP:Weitenberg2011} or, by using alkaline-earth(-like) atoms, the $^{1}S_0$ ground 
 states and the very long-lived $^{3}P_0$ excited states can be manipulated completely independently 
 by laser light, allowing the construction of independent optical lattices for the two states~\cite{Daley2008}.}

\ann{
Finally, we underscore that the quantum computing models discussed above for optical lattice 
implementations require state-dependent moving potentials and that the entanglement among 
qubits is produced by means of collisions between ultracold atoms . However, we would like to 
mention that, alternatively, the entanglement between trapped atoms in optical lattices can be 
achieved through the laser-induced coherent electric dipole-dipole interaction of pairs of atoms 
that are made to occupy the same well~\cite{QIP:Brennen99}. Such an approach has two main 
advantages: neutral atoms are weakly coupled to the external environment, thus the detrimental 
effects of decoherence are strongly suppressed, and operations can be performed in parallel on 
a large atomic ensemble, which enables massive entanglement production.}


\an{
\section{Other quantum computation models}}

\subsection{Cavity QED based schemes}
\label{QIP:sec:cavityQED}

Recent experimental advances in cavity QED on a chip have yielded results 
that promise the full integration and scalability of such
cavities.
Microscopic Fabry-Perot cavities whose open structure gives access 
to the central part of the cavity field have been developed. In such cavities 
strong coupling between a single atom and the cavity mode has been obtained \cite{QIP:Colombe07}.
In these experiments a BEC was employed, which can be located deterministically
everywhere in the cavity and positioned entirely within a single antinode 
of the standing-wave cavity mode field. This gives rise to a 
controlled and tunable coupling rate. 

On the theoretical side, proposals for a quantum computer based on a cavity
QED model have been put forward. The scheme of Ref.~\cite{QIP:Pellizzari95} assumes 
$N$ atoms coupled to a single quantized mode of a high finesse cavity. Quantum 
operations (e.g., a controlled-NOT gate) are realised via the coupling of the atoms with individual lasers 
and their entanglement is mediated by the exchange of a single cavity photon.  
In a similar setup a controlled-NOT gate can be performed through 
a sequence of (destructive) measurements on the atoms and quantum 
non-demolition measurements of the atom number \cite{QIP:Sorensen03}. 
Because of the randomness of the measurement outcome in this case the gate
operation is probabilistic. Nevertheless, this scheme is more robust against 
decoherence and cavity losses than the one of Ref.~\cite{QIP:Pellizzari95}: 
while in that proposal the infidelity scales as $1/\sqrt{2C_0}$, 
in Ref.~\cite{QIP:Sorensen03} it scales as $\log(2 C_0)/2C_0$, where $C_0$ 
is the cooperativity parameter.

\subsection{Rydberg and ensemble based schemes}

Based on such experimental achievements, schemes that exploit the 
interaction of atoms excited to low-lying Rydberg states~\cite{QIP:Jaksch00,Brion2007a,Brion2007b} 
\an{are also an appealing solution for QIP with neutral particles}. An
appropriate sequence of laser pulses above a waveguide can excite the qubits 
into Rydberg states and entangle them via electric dipole-dipole 
interactions \cite{QIP:Christandl04} \lmn{and by triggering
electromagnetically induced transparency in the atomic ensemble 
it is also possible to transfer atoms between 
internal states~\cite{Mueller2009}}. While the 
phase gate model suggested in Ref.~\cite{QIP:Jaksch00} relies only on the 
strong dipole-dipole interaction, in another 
recent proposal \cite{QIP:Moller08} one can combine the Rydberg blockade 
mechanism\lmn{, recently observed for two atoms~\cite{Urban2009,Alpha2009},} 
with the rapid laser pulse sequence of the well-known stimulated Raman 
adiabatic passage (STIRAP). This combination with the engineering of a 
time-dependent relative phase $\phi_{\rm R}(t)$ between the Rabi frequencies 
of the two STIRAP laser pulses affords a higher degree of control of the 
phase of Eq.~(\ref{eq:ideal-mapping}) through the manipulation of geometrical phases. 
We briefly mention that recently the first demonstration of a CNOT gate between 
two individually addressed neutral atoms by means of the Rydberg blockade interaction 
with an experimental gate fidelity of about 0.72 and $\pi$ pulse times of $\sim 750$ ns 
has been demonstrated \cite{QIP:Isenhower10}. 

\lmn{A very recent theoretical analysis~\cite{Goerz2011} shows that not only 
the interaction strength among the qubits is a limiting factor for the gate operation time, 
but for resonant excitation, the interaction between two atoms causes a coupling between 
electronic and nuclear dynamics. This induces vibrational excitation which can be carried away 
by the laser pulse only if the target state is fully resolved. Hence, the excitation of atoms into Rydberg
states yields an interaction that one might expect to allow for nanosecond to sub-nanosecond 
gate operation times~\cite{QIP:Jaksch00}, but since the motional state of the atoms needs to be 
restored at the end of the gate, gates with sub-nanosecond operation time seem difficult to realize.}

So far, we discussed only qubits that are represented by individual two-level
systems. This kind of qubits requires high control and addressability of 
individual particles, which raise important challenges for experimentalists. 
Some theoretical investigations, however, use a symmetric collective state 
of a mesoscopic atomic ensemble \cite{QIP:Lukin01}, where only one excitation 
is present in the system. The dipole excitation blockade mechanism of Rydberg 
atoms works in that direction, because it prevents multiple excitations 
in an ensemble. The same mechanism used to implement a fast phase gate 
with two Rydberg atoms can be extended to qubits stored in few-$\mu$m-spaced 
atomic clouds, where each atomic ensemble is a qubit. Alternatively, collective states 
in ensembles of multilevel quantum systems can be used to store the
information \cite{QIP:Brion07}. Quantum operations such as one- and two-bit gates are then 
implemented by collective internal state transitions taking place in the
presence of an excitation blockade mechanism, such as the one provided by the
Rydberg blockade. 
\lmn{For more details on QIP with Rydberg atoms we refer to Ref.~\cite{Saffman2010}.}

\subsection{\ann{Polar molecules based schemes}}
\label{sec:polar}

\ann{
Up to now, we discussed schemes based on the manipulation of single or ensembles of 
trapped atoms, but a quite interesting solution to the construction of a quantum computer of large scale 
is given by the use of polar molecules \cite{QIP:DeMille02}. Indeed, they incorporate two 
important features of neutral atoms and ions: long coherence times and strong 
interactions. Besides this, polar molecules have stable internal states that can be controlled 
by electrostatic fields. This controllability is due to their rotational degree 
of freedom in combination with the asymmetry of their structure. 
By applying moderate laboratory electric fields, rotational
states with transition frequencies in the microwave range can be mixed, 
and the molecules acquire large dipole moments, which are the key
property that makes them effective qubits in a quantum
processing system. Furthermore, the application of electric field
gradients leads to large mechanical forces, allowing to trap the
molecules. For instance, the electrostatic Z-trap for polar molecules proposed
in Ref.~\cite{QIP:Andre06} creates a non-zero electric 
field minimum in close proximity to the surface, analogous to 
Ioffe-Pritchard type magnetic traps for neutral atoms.
}

\ann{
In the pioneer proposal by DeMille \cite{QIP:DeMille02}, qubits are represented by rotational states 
of a diatomic polar molecule and the coupling among qubits is realized by means of the electric 
dipole-dipole interaction, which can be also produced by using Raman transitions between 
scattering and bound states of heteronuclear molecules trapped in optical lattices~\cite{QIP:Lee05}.  
The integration of a 1D optical lattice with a superimposed crossed dipole trap allows to confine polar 
molecules in sites spaced by $\lambda/2$. Such a scheme allows to confine in a trap of length 5 mm about 
$10^4$ qubits for a laser wavelength $\lambda\sim 1\mu\mathrm{m}$ and beam waist of 50 $\mu$m 
with moderate decoherence rates. Without an external electric field, however, polar molecules 
have no net dipole moment, and therefore the application of an external electric field is required, which 
mixes the rotational states. For weak fields the mixing state that arises from the $J=0$ ($J=1$, 
$m_J=0$) state corresponds to a dipolar charge distribution along (against) the external homogeneous 
electric field. The logical state $\ket{0}$ ($\ket{1}$) corresponds to the dipole moment oriented along (against) 
the external electric field. Moroeover, the addition of an electric field gradient allows the addressing of 
individual molecules. By a proper choice of electric field strengths it is possible to realize, with KCs 
molecules, a CNOT gate in a time of about 50 $\mu$s, thereby enabling about $10^5$ quantum 
operations within the decoherence time of about 5 s, whose main source is photon scattering from the 
trap laser \cite{QIP:DeMille02}. The readout of the computation can be accomplished by state-selective, 
resonant multiphoton ionization and imaging detection of the resulting ions and electrons or, alternatively, 
with nearby single-electron transistors~\cite{QIP:Schoelkopf98}, which allow to detect the molecular dipole fields.}

\ann{The above outlined estimates are quite encouraging, but the analysis does not take into account, for instance, 
the effects caused by the motional states of the molecules, which may induce an additional gate fidelity loss, or the hyperfine 
structure of the molecules, which makes the initialization of the quantum computation and the operation of quantum 
gates more complicated.  Another source of decoherence for such a scheme arises from the interaction 
of non-nearest-neighbor qubits, which cannot be switched off locally. The negative effects of such interaction 
can be removed by using a refocusing (spin-echo) technique, similar to the one adopted in nuclear magnetic 
resonance quantum computation~\cite{QIP:Nielsen00}, or, alternatively, by using heteronuclear molecules, 
which have a different dipole moments for different electronic, vibrational or rotational states and zero 
expectation value for the dipole moment in the $N=0$ rotational state ($N$ represents the quantum 
number of the angular momentum $\hat{\bf{N}}$ of the nuclei of the molecule). This approach allows to make 
the dipole-dipole interaction ``switchable", therefore making the experimental realization of two-qubit quantum 
gates less complicated as well as minimizing decoherence. With this method it has been shown that up to $10^6$ 
gate operations are obtainable within the coherence times, for instance, of NaCl or CaF molecules~\cite{QIP:Yelin06}. 
Another way to switch on and off the dipole interaction is given by the use of Raman transitions, which transfer
the qubit state from a set of non-interacting levels, used for the storage of information, to a set of interacting levels, 
which allow the qubits to interact. Entanglement can then be yielded by simple optical pulses of relatively short duration 
(i.e., larger than 200 ps) thus allowing for relatively short gate operation times ($\simeq 10\,\mu\mathrm{s} - 1\,\mathrm{ms}$)
~\cite{QIP:Charron07}. Instead, by using vibrationally excited (non-polar) molecules, like acetylene ($\mathrm{C}_2\mathrm{H}_2$), 
entanglement of two qubits via molecular bindings can be (theoretically) performed, by using optimal shaped femtosecond laser 
pulses in the infrared regime, even on the picosecond timescale~\cite{QIP:Tesch02}. 
}

\subsection{\an{Hybrid QIP implementations}}
\label{sec:hybrid}

\gan{
In this section we briefly describe some hybrid QIP schemes, namely combinations of solid-state and atomic qubits 
in a single quantum processor, as we outlined in the introduction of the paper. }

\gan{The proposals of Refs.~\cite{QIP:Sorensen04,QIP:Tian04b,QIP:Andre06,QIP:Rabl06,QIP:Tordrup08,QIP:Verdu09} 
suggest to use superconducting wires or superconducting microwave cavities to `wire up' several atomic qubits or 
to couple atomic qubits to a Cooper pair box on the chip surface, as we briefly described in the introduction. Most of the 
schemes consider polar molecules \cite{QIP:Andre06,QIP:Rabl06,QIP:Tordrup08a,QIP:Tordrup08} as atomic 
system, but other quantum memories can be realized by Rydberg atoms \cite{QIP:Sorensen04}, atomic ions 
\cite{QIP:Tian04b}, ground state neutral atoms \cite{QIP:Verdu09}, and electron spin ensembles \cite{QIP:Wesenberg09}.
Hereafter we discuss only the schemes based on polar molecules.}

\gan{
The proposals of Refs.\cite{QIP:Rabl06,QIP:Tordrup08} suggest a quantum computer 
model where ensembles of cold molecules are used as a stable quantum memory 
by means of collective spin states, whereas a Cooper pair box, connected to the molecular 
ensemble via a stripline cavity, is used to perform one- and two-qubit operations and 
readout (see also Fig.~\ref{QIP:fig:Hquantumprocessor}). In Ref.~\cite{QIP:Rabl06} the 
ground state and a symmetric collective state with only one excitation present in the ensemble 
are the qubit states, and therefore an ensemble of molecules carries only 
two logic states. 
On the other hand, Ref.~\cite{QIP:Tordrup08} proposes a `holographic' memory 
consisting of $N$ molecules in a lattice, initially all in the same internal quantum 
state $\ket{g}$ [see Fig.\ \ref{QIP:fig:holo}(b)]. 
\begin{figure}
\includegraphics[width=8cm]{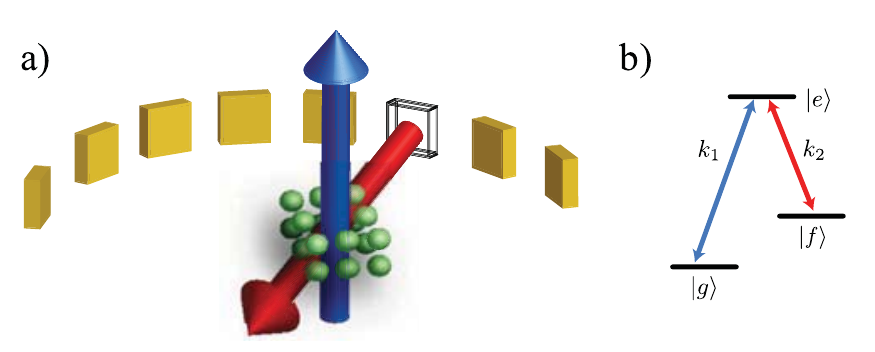}
\caption{\gan{(Color online). (a) By varying the direction of a
control field $\Omega_2(t)e^{i{\bf k}_2\cdot{\bf x}}$, an
incident single photon with wave vector ${\bf k}_1$ may be
transferred to different collective storage modes with wave
vector ${\bf q}={\bf k}_1-{\bf k}_2$. (b) The levels
$\ket{g}$ and $\ket{f}$ are coupled by a two photon process
leaving no population in the electronically excited state
$\ket{e}$ (adapted from Ref.~\cite{QIP:Tordrup08}). 
Copyright (2008) by The American Physical Society.}}
\label{QIP:fig:holo}
\end{figure}
In this case the quantum information in an
incident weak field $\Omega_1e^{i{\bf k}_1\cdot{\bf x}}$, 
by means of a control field
$\Omega_2(t)e^{i{\bf k}_2\cdot{\bf x}}$ and the Hamiltonian
\begin{equation}\label{eq:Hk}
\hat H_{{\bf q}}=\sum_{j=1}^N
\Omega_1e^{i{\bf k}_1\cdot{\bf x}_j}\ket{e}_{jj}\bra{g} +
\Omega_2e^{i{\bf k}_2\cdot{\bf x}_j}\ket{e}_{jj}\bra{f} +
\text{h.c.},
\end{equation}
is transferred onto a collective matter-light excitation which
propagates slowly through the medium and is brought to a
complete stop by turning off $\Omega_2(t)$.}

\gan{
The coupling in Eq.\ (\ref{eq:Hk}) can be used to map a
single-photon state to the collective phase pattern state
$\ket{f,{\bf q}} \equiv 1/\sqrt{N} \sum_j e^{i{\bf q}\cdot
{\bf x}_j} \ket{g_1 \ldots f_j \ldots g_N}$, where
${\bf q}={\bf k}_1-{\bf k}_2$ is the wave number difference
of the two fields. The set of collective excitations 
$\{\ket{f,{\bf q}_s}: s=1,\dots,K\}$ can be used to simultaneously
encode up to hundreds of qubits in just one sample by associating the logical
state $\ket{b_1 b_2 \ldots b_K}$ ($b_i=0,1$) with the
collective state $\ket{f,{\bf q}}$. Addressing
different qubits is then accomplished by applying laser
beams from different directions such that the orthogonality condition 
$\bra{f,{\bf q}_s}f,{\bf q}_j\rangle$ is approximately fulfilled 
(i.e., $ \approx \delta_{{\bf q}_s,{\bf q}_j}$), as illustrated in Fig.~\ref{QIP:fig:holo}(a). This way of 
encoding does not pose special problems for atom-chip-like devices, 
where usually single-atom addressing with laser beams is required. }

Finally, we briefly mention, that the hybrid system ``atom-molecule" also provides an interesting 
solution for fast and robust quantum computation~\cite{QIP:Kuznetsova10}. An optical lattice with 
two different atomic species per site, like $^{87}$Rb and $^7$Li, forms the quantum register. The 
quantum information is again stored in hyperfine levels of one neutral atom while the other 
atom is in a state stable against collisions, such that at ultralow temperatures the scattering of the two 
atomic species is elastic, thus preventing atom loss and qubit decoherence. Two-qubit operations, like 
the phase gate, are then accomplished by converting pairs of atoms at two lattice sites into stable 
molecules with a large dipole moment. The produced molecules can then interact via strong 
dipole-dipole interactions, therefore resulting in fast quantum gates as discussed in Sec.~\ref{sec:polar}. 
Such a system can be also utilized for investigating the generation of many-body entanglement and 
new Hubbard Hamiltonians~\cite{QIP:Trefzger09}. 
\lmn{Alternatively, one can use the interaction between a Rydberg atom and a polar molecule in order 
to realize quantum operations between two molecular qubits~\cite{QIP:Kuznetsova11}. A polar molecule within the electron 
state in a Rydberg atom can either shift the Rydberg state or produce a Rydberg molecule in a state-dependent 
manner, resulting in molecular state dependent van der Waals or dipole-dipole interaction between Rydberg atoms.
}


\section{Conclusions and Outlook}

We have reviewed a broad range of theoretical proposals for implementing quantum gates 
with \an{neutral particles and how they can be realized on a chip.} Several of these proposals 
have been worked out in detail, including investigations of various kinds of imperfections 
such as decoherence due to atom-surface interactions. High-fidelity quantum gates compatible 
with the requirements for fault-tolerant QIP seem experimentally feasible even 
in the presence of these imperfections. 

\tunnu{On the experimental side, impressive progress was made in the chip-based coherent control 
of ultracold atoms. Coherent manipulation of long-lived hyperfine qubit states was demonstrated 
close to a chip surface \cite{QIP:Treutlein04}, and coherent control of motional states is routinely 
achieved in chip-based atom interferometers \cite{QIP:Wang05,QIP:Hofferberth06,QIP:Boehi09}. 
Chip-based lattices were created \cite{QIP:Whitlock09} which could store a large register of qubits. 
The experimental achievement of single-atom preparation and detection on an atom chip with a fidelity 
exceeding 99.92\% \cite{QIP:Gehr10} represent important milestones on the way to atom chip based QIP. 
Another important milestone is the on-chip manipulation of ultracold atoms with an internal-state 
dependent potential \cite{QIP:Boehi09}, a key ingredient of collisional quantum gates. }
In addition to this, the recent achievement of the fluorescence imaging of strongly 
interacting bosonic Mott insulators in an optical lattice with single-atom and single-site resolution 
\cite{QIP:Sherson10} strengthens the possibility to effectively realize optical lattice based 
QIP schemes, as we discussed in Sec. \ref{sec:optical}. In optical lattice experiments the 
fidelity for the identification of atoms at a given lattice site is about 98\%~\cite{QIP:Bakr09}. 

\tunnu{With these results, all individual elements for the chip-based two-qubit gate of 
Refs.~\cite{QIP:Calarco00a,QIP:Treutlein06b} have now been demonstrated experimentally. 
An important challenge for the near future is to combine them in a single experiment. In the 
related context of quantum metrology with atomic ensembles, a recent experiment has already 
demonstrated the generation of entangled atomic states on an atom chip \cite{QIP:Riedel10}, 
using techniques similar to the ones proposed for the quantum gate of Refs.~\cite{QIP:Calarco00a,QIP:Treutlein06b}.}  

While the development of chip-based near-field traps was pioneered for ultracold neutral atoms, it 
has already triggered similar developments for other systems such as ions or molecules. As trapped 
ions are currently one of the frontrunners in the field of QIP, the recent demonstration of chip-based 
ion traps is particularly promising \cite{QIP:Herskind10,QIP:Stick06,QIP:Schulz06,QIP:Seidelin06} as 
well as the strong coupling between an ion Coulomb crystal and a single mode cavity field \cite{QIP:Herskind09}.

We would like to conclude by briefly mentioning one of the most promising directions of future research 
for QIP. Indeed, a particularly attractive feature of chip traps is the possibility to combine atomic or molecular qubits with 
solid-state qubits on the chip surface \cite{QIP:Sorensen04,QIP:Andre06,QIP:Rabl06,QIP:Tordrup08,QIP:Verdu09}. 
Such hybrid systems would combine fast processing in the solid-state with long coherence times for information 
storage in the atomic system. An impressive degree of coherent control has been demonstrated e.g.\ for qubits 
based on superconducting circuits \cite{QIP:Steffen06,QIP:Majer07,QIP:Kubo10,QIP:Wu10}. Coherent dynamics as well as decoherence 
in these systems typically occur on a time scale of nanoseconds to microseconds, several orders of magnitude 
faster than in atomic gases. A very promising approach to couple atomic and superconducting qubits is the use 
of superconducting microwave resonators \cite{QIP:Andre06,QIP:Rabl06,QIP:Tordrup08,QIP:Verdu09}.
To combine the necessary cryogenic technology with atom chips represents an experimental challenge 
which is currently pursued in several experiments.


\section*{Acknowledgements}
\toni{
We acknowledge financial support by  the IP-AQUTE (P.T.,T.C.), 
SFB/TRR21 (A.N.,T.C.), the Marie Curie program
of the European Commission (Proposal No. 236073,
OPTIQUOS) within the 7th European Community Framework
Programme and the 
Forschungsbonus of the University of Ulm and of the UUG (A.N.), 
and the Swiss National Science Foundation (P.T.). }


\bibliography{atomchips}

\end{document}